%% file: neu2.tex
\documentclass[11pt,
]{article} 
\usepackage[utf8]{inputenc}

\usepackage{amsmath}
\usepackage{amssymb}
\usepackage{amsthm} 
\usepackage{ytableau}
\usepackage{multirow}
\usepackage{array}
\usepackage{float}

\usepackage{geometry} 
\usepackage{graphicx} 
\usepackage{ytableau}
\usepackage{tikz}
\usetikzlibrary{decorations.pathreplacing}
\usepackage{epstopdf}
\DeclareGraphicsExtensions{.eps,.pdf,.png}
\input xy
\xyoption{all}

\input{SDefin.tex}

\newcommand \np {*(blue!8)5}
\newcommand \tsy \textstyle

\usepackage{capt-of}
\usepackage{mathtools} 
\usepackage{subcaption}
\usepackage{booktabs}

\usepackage{vmargin} 
\setpapersize{A4}
\setmargins
		{1.1cm}
		{0.7cm}
         	{19.0cm}
         	{24.5cm}
         	{12pt}
         	{1cm}
         	{0pt}
         	{1.25cm}

\title{On financial applications of the two-parameter \\ \PD distribution\\
 {\em \small \cldb \sc  Research Note}
}
\author{  Sergey Sosnovskiy 
 \\\tt \small ssnv.sky@gmail.com
}
\small \date{July 7, 2015} 

\begin{document}
\maketitle
\abstract{ 


Capital distribution curve is defined as log-log plot of normalized stock capitalizations ranked in descending order. The curve displays remarkable stability over periods of time. 

Theory of exchangeable distributions on set partitions, developed for purposes of mathematical genetics and recently applied in non-parametric Bayesian statistics, provides probabilistic-combinatorial approach for analysis and modeling of the capital distribution curve.
Framework of the two-parameter Poisson-Dirichlet distribution contains rich set of methods and tools, including infinite-dimensional diffusion process.

The purpose of this note is to introduce framework of exchangeable distributions on partitions in the financial context. In particular, it is shown that averaged samples from the Poisson-Dirichlet distribution provide approximation to the capital distribution curves in equity markets. This suggests that the two-parameter model can be employed for modelling evolution of market weights and prices fluctuating in stochastic equilibrium.



}

\newpage
\section{Introduction}

The {\em capital distribution curve} is defined as log-log plot of stock market weights ranked in descending order. Temporal stability of the shape of this curve is one of the cornerstones of the
Stochastic Portfolio Theory (SPT), developed by Fernholz, Karatzas et al.  (\cite{fernholz2002stochastic}, \cite{karatzas2009stochastic} and \cite{fernholz2013second}). In contrast to the MPT and the CAPM, which are based on normative assumptions, the Stochastic Portfolio Theory is a {\em descriptive} theory, since it studies empirical dynamics and characteristics of equity markets. In particular, the SPT captures tendency of stocks of retaining their ranks.
 The SPT model employs machinery of rank-interacting Brownian particles and semimartingales. 

Framework of partition structures, imported from mathematical genetics and comprising combinatorial and probabilistic methods,  
provides complementary approach for modeling and analysis of the capital distribution curve and can be summarized as follows.
\begin{itemize}
\item The market is considered as a large combinatorial structure - partition of the set of  the invested units of money. Capitalizations of individual stocks correspond to block or cluster sizes of the partition, represented by integers, for instance, measured in cents. 
\item Number of set partitions defines number of ways each partition can be realized combinatorially. In other words, the market can be represented as a giant Young diagram with vector of capitalizations determining (potentially very large) number of ways such market configuration can be realized.
\end{itemize}
\vskip -0.1cm
Partition structures are important for several reasons.
\begin{itemize}
\item First of all, partition structures provide a model of random transitions with dynamic dimensions. In other words,  at any time number of diffusion components may change due to appearance of a new stock or bankruptcy of existing firm. 
\item Second, partition structure, with non-trivial limiting distribution, defines {\em asymptotic shape of the corresponding combinatorial structure}. 
In particular, mechanism of  shape formation provides an explanation of the phenomenon of stability of the capital distribution curve.
\end{itemize}

The \tpPD model is a remarkable and well studied instance of partition structures. It possesses analytically tractable limiting distribution defined in the simplex of ranked weights. 
\vskip 0.1cm
{\cldb \bf \PD distribution.}
 The Dirichlet distribution with $m$-dimensional vector of parameters $(\al_1,...,\al_m)$ defines probability for non-negative proportions in a standard simplex. 
Kingman \cite{kingman1975random} considered limiting behavior  of this distribution with symmetric vector of parameters $(\al,...,\al)$ such that $\te=m\al=const$ for $m\to\infty$ and called distribution of ranked components the Poisson-Dirichlet ($\mathcal{PD}$) distribution (with one parameter $\te$). 
This distribution is defined in the infinite simplex of ranked weights, known as Kingman simplex 
\[ \nabla =\big\{x_1 \ge x_2 \ge ...\, \big | \, x_i\ge0, \textstyle\sum x_i=1 \}\]
Size-biased permutation provides an efficient method of sampling from the Dirichlet and the Poisson-Dirichlet distributions.  In a framework of population biology Engen \cite{engen1978} suggested modification of the size-biased method, which produced another class of \PD distributions. It was called the \tpPD distribution by Perman, Pitman and Yor, who   rediscovered it  in the context of studying of ranked jumps of gamma and stable subordinators (see \cite{perman1992size},\cite{PY}). 
Monograph by Pitman \cite{pitman2002combinatorial} contains wealth of information on   the \tpPD model.
As shown by Chatterjee and Pal \cite{chatterjee2010phase}, limiting behaviour of rank-interacting system of Brownian particles is characterized by the $\mathcal{PD}(\al,0)$ distribution.

Aoki pioneered applications of \exc distributions in economics (\cite{aoki2001modeling},\cite{Aoki228}), in particular using finitary characterization by  Garibaldi, Costantini,  et al. (\cite{garibaldi2004finitary}, see also book \cite{garibaldi2010finitary}). Markov chain approach with transitions in space of partitions was independently developed by Garibaldi, Costantini,  et al. \cite{garibaldi2004finitary}, \cite{garibaldi2007two}.
Petrov \cite{petrov2009two}, inspired by works of Kerov, Fulman \cite{fulman2005stein},  Borodin and Olshanski \cite{borodin2009infinite} constructed a diffusion process preserving the \tpPD distribution in the infinite-dimensional ranked simplex. 

\vskip 0.15cm
This research note aims at illustration of applications of partition structures and the two-parameter model for modeling of stochastic evolution of the capital distribution curve. In particular, it is shown in Section \ref{sec-examples} that the two-parameter model provides reasonable approximation of capital distribution curves in equity markets. Moreover the model also provides fit for distribution of relative total capitalizations of stock exchanges.

Main results of this paper were presented at the 
8th World Congress of the Bachelier Finance Society, 2014. 
{The author is very grateful to Prof. I. Karatzas for useful advice and suggestions.}

\newpage
\subsection{Capital distribution curve} \label{susec-cdc}

Log-log plot of ranked market weights 
displays
\begin{itemize}
	\item power law behavior,
	\item concavity of the curve and
	\item stability over periods of time
\end{itemize}
For example, figure below shows capital distribution curves of the NASDAQ market on three dates in 2014. 
\footnote{Data source is {\tt http://www.google.com/finance\#stockscreener}}
As it can be  seen from the chart most of market weights had relatively small fluctuations, despite significant fluctuations of NASDAQ market capitalization during that period of time.
Stability of the capital distribution curve suggests certain {\cldb \em 
independence of market weights and overall market capitalization}.
\begin{figure}[H]
\centering
\includegraphics[width=0.7\linewidth,
trim=2.1cm 10.5cm 2cm 11.75cm,clip]{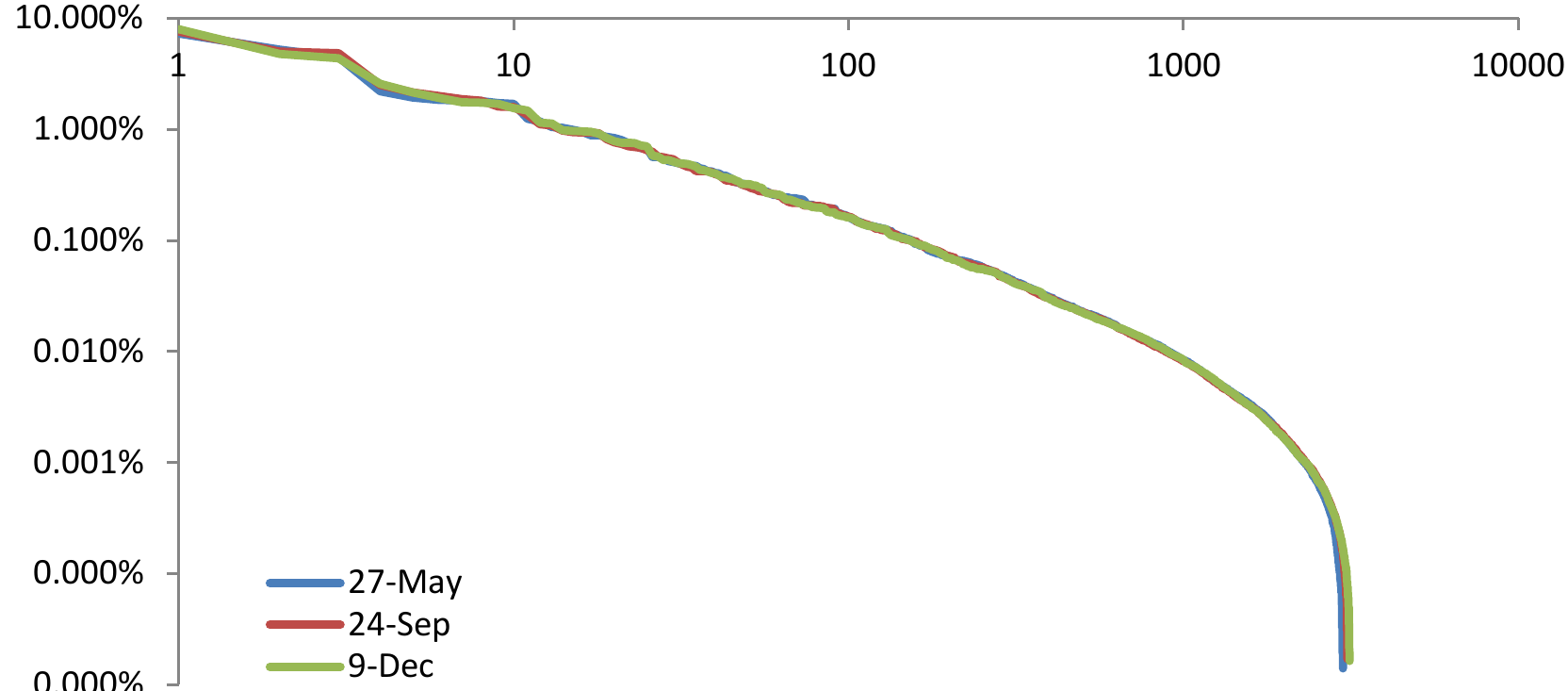}	
\caption{\small NASDAQ, capital distribution curves on May 27, Sep 24, Dec 9, 2014  }\label{fig2}
\label{fig2}
\end{figure}
More detailed chart reveals behavior of weights of top 100 stocks. 
\begin{figure}[H]
\centering
\includegraphics[width=0.7\linewidth,
trim=2.1cm 10.25cm 2cm 11.75cm,clip]{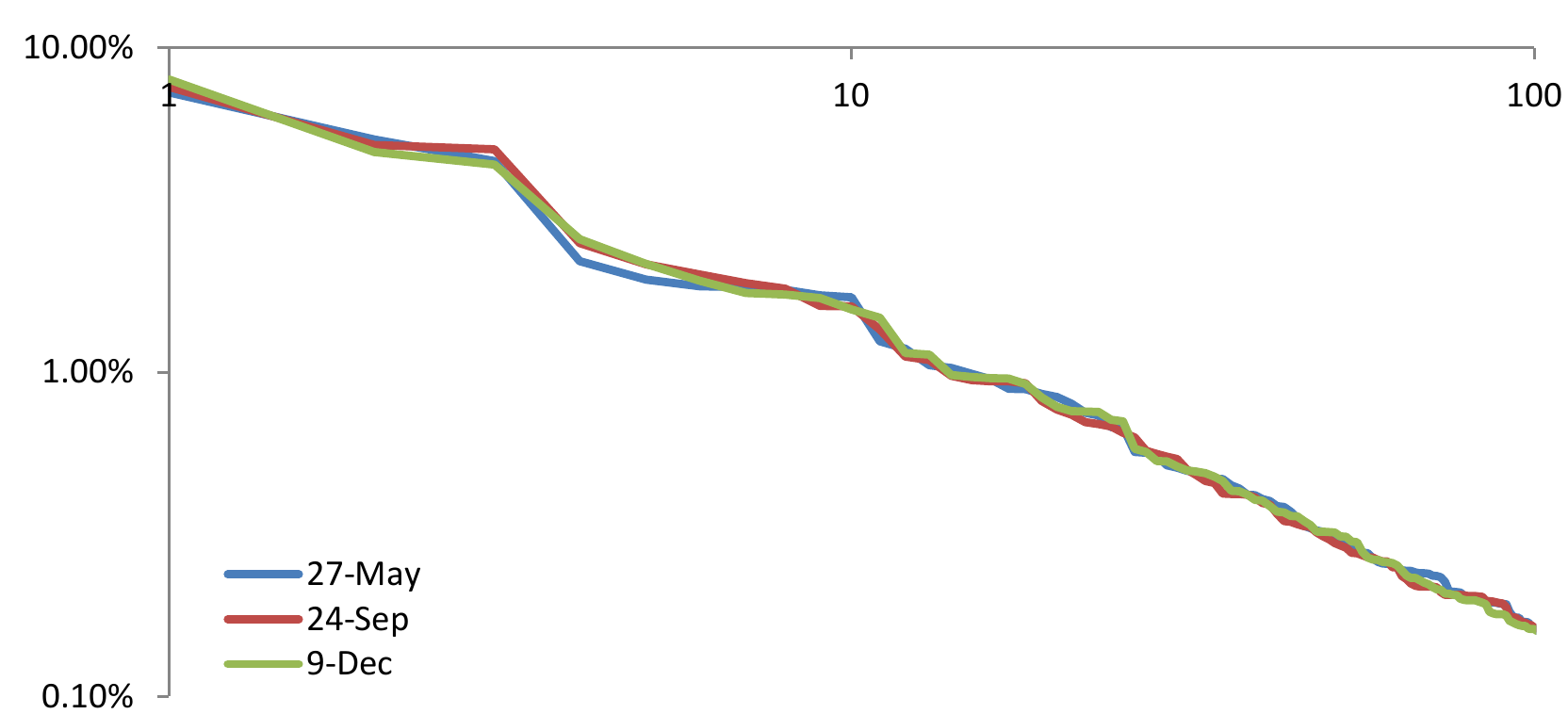}
\caption{\small weights of top 100 stocks, NASDAQ}\label{fig3}
\label{fig3}
\end{figure}
Capital distribution curves on majority of equity markets, as well as distribution of capitalizations of world stock exchanges, have shapes similar to one shown at Figure \ref{fig2}. Section \ref{sec-examples} contains examples of fit of these curves by the \pd-model.

\newpage
\subsection{\PD distribution and market weights}
Log-log plot of ranked samples from the \PD law is characterized by
\begin{itemize}
	\item power law behavior,
	\item concavity of the curve and
	\item stability around average shape
\end{itemize}
The  infinite-dimensional \PD distribution generalizes symmetric finite-dimensional Dirichlet distribution.
Moreover, as shown in Section \ref{represent}, both distributions can be represented by normalization of sequences of 
random variables $(y_1, y_2, ...)$ by their sum $S=\sum y_j$ 
\[( {y_1}/s,\;  {y_2}/s, \;\dots )\]
{\cldb with the property of independence of weights and the sum $S$}.

Figure below illustrates fit of NASDAQ market weights by averages of samples from the two-parameter distribution. Estimation of parameters is by least squares method.

\begin{figure}[H]
\centering
  \includegraphics[width=14cm, height=7cm,
	trim=1.5cm 7cm 2cm 6.7cm,clip]{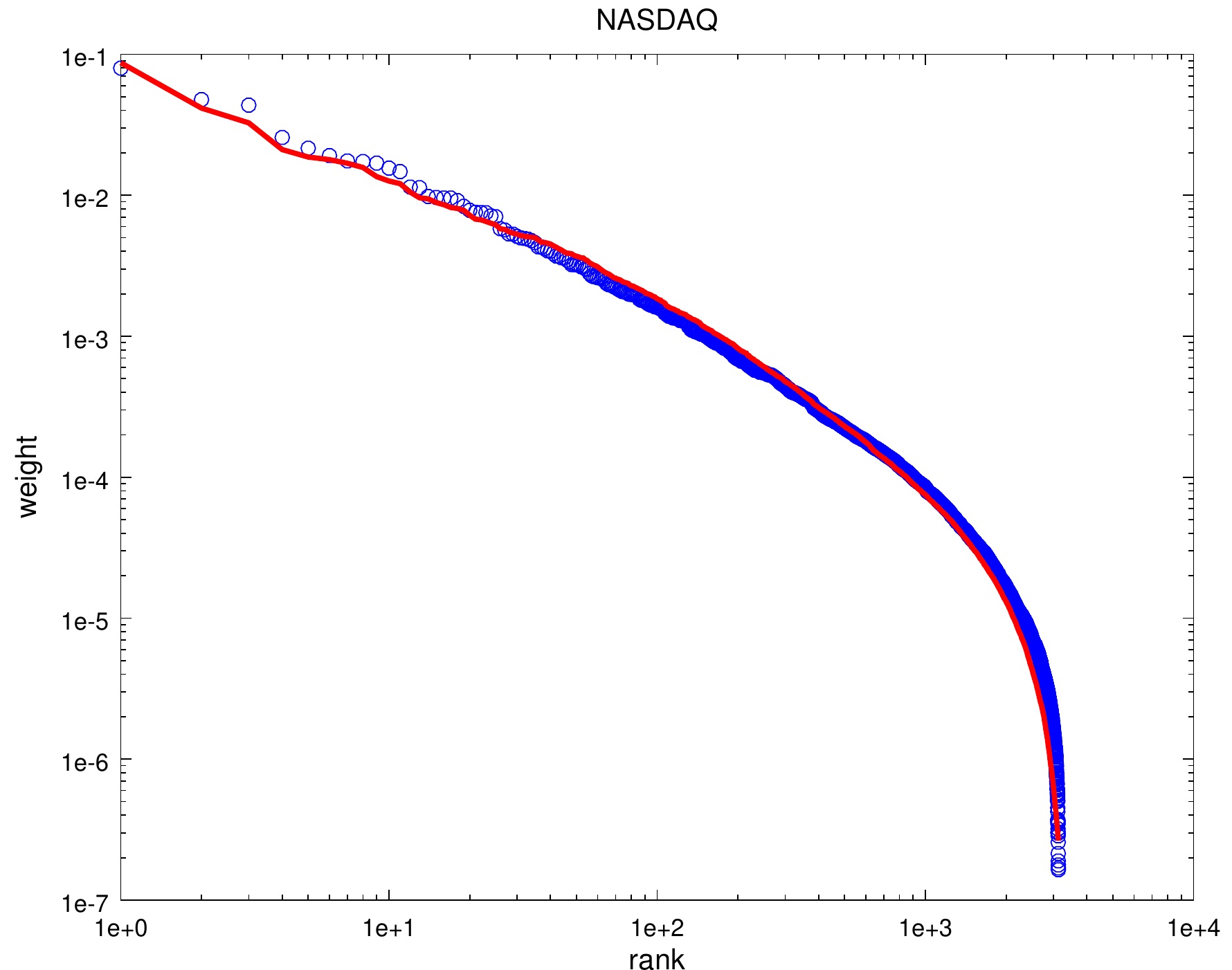}
  \caption{\small NASDAQ fit by $\mathcal{PD}(0.60,55 )$,   (data as of Dec 9, 2014)}
  \label{fig4}
\end{figure}
Next figure displays typical behaviour of ranked random weights 
\begin{figure}[H]
\centering
  \includegraphics[width=14cm, height=7cm,
	trim=1.5cm 7cm 2cm 6.7cm,clip]{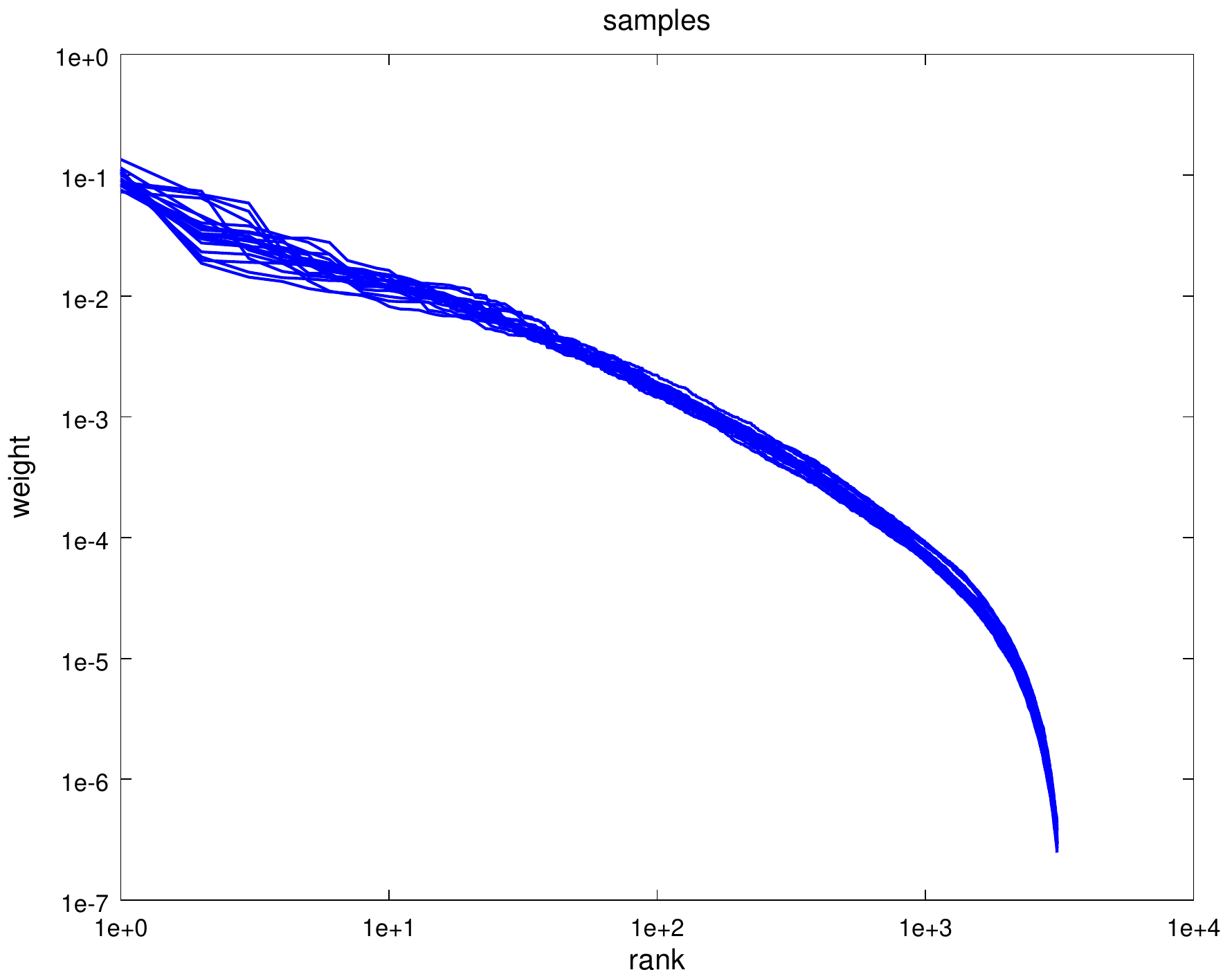}
   \caption{\small 20 sample paths of $\mathcal{PD}(0.60, 55)$}
  \label{fig5}
\end{figure}

\subsection{Ranked capitalizations and market weights}
Stock capitalization at time $t$ is calculated as product of the shares outstanding and the stock price \[C_n(t)=q_n(t) \cdot P_n(t)\]
For capitalizations ordered as
$
	C_{(1)}(t)\ge C_{(2)}(t)\ge \dots
$ 
corresponding ranked market weights are determined by
\vskip -2mm
\[x_{(n)}(t)=\frac{C_{(n)}(t)}{M(t)},\]
where $M(t)=\sum C_n(t)$ is total market capitalization at time $t$. 
Stability of the capital distribution curve means 
\vskip -2mm
\[
	 x_{(i)}(t)\approx\EE x_{(i)}(t+\Delta t)
\] 
In other words, ranked weights remain approximately the same despite changes in capitalizations. This implies that for relatively short periods of time, when the stock retains its rank 
{\em \nr}approach of pricing approximately holds
%
\[	
	 \frac{P_{(n)}(t)}{M(t)} \approx \EE \;
	  \frac{P_{(n)}(t+\Delta t)}{M(t+\Delta t)}
\]
However, it should be noticed that 
the longer the time period $\Delta t$, the less likely that stock retains its rank. More advanced approach of modelling market weights and stock capitalizations is based on application of diffusion theory and representation of the $\PDat$ distribution in terms of jumps of subordinators. This representation is  
known as Proposition 21 in the celebrated paper of Pitman and Yor \cite{PY}.

\subsection{Gamma-Dirichlet algebra}\label{represent}
There is close relationship between the gamma and Dirichlet distributions, characterized by number of important properties, 
which in the symmetric case can be summarized as follows.
Let us consider $m$ independent and identically distributed gamma variables 
$y_i \sim \Gd(\al,c)$ with shape $\al$ and scale $c$. 
The first, {\em convolution} property states that the sum of these variables $S=\sum y_i$ also has gamma distribution $S \sim \Gd(\te,c)$ with $\te=m \al$. 
The second property states that 
normalized components $x_i=y_i/S$ are {\em independent} of the sum $S$, moreover,  as it  has been shown by Lukacs \cite{lukacs1955characterization}, this characterizing property holds 
if and only if $y_i$ are gamma distributed with the same scale $c$. 
Finally,  
normalized vector $\vc x =\big({y_1}/S,...,{y_m}/S\big)$ has symmetric Dirichlet distribution $\vc x \sim \Dir_m(\al)$. 

Conversely, with Dirichlet distributed vector $\vc x \sim \Dir_m(\al)$ and independent gamma distributed $S \sim \Gd(\te,c)$ 
'restored' variables $y_i=x_i \cdot S$, correspondingly, have gamma distributions 
$y_i \sim \Gd(\al,c)$. 

Obviously, these properties hold as well in the case of the ordered Dirichlet distribution. For instance, with ranked components  
 $x_{(1)}\ge ... \ge x_{(m)}$ obtained from the symmetric Dirichlet distribution and independent $S \sim \Gd(\te,c)$, restored gamma variables $y_{(m)}=x_{(m)} \cdot S $  are also ranked in descending order.

Similar characterization of the $\PDat$ law 
is provided by the Proposition 21 
in Pitman and Yor \cite{PY}, 
which informally can be restated as follows.
Let us consider tempered stable subordinator $f_t$ with \Lv density 
$\textstyle\nu(y)=\frac{ \al}{\GG(1-\al)}y^{-\al-1}e^{-y}$ in random time interval $[0,T]$, with $T\sim \Gd(\te/\al,1)$ and denote ranked jumps of the subordinator in this interval by $\eta_{(1)} \ge \eta_{(2)} \ge \dots$. 
Sum of these jumps is equal to value of the tempered subordinator stopped at random time $T$ 
\[S=\textstyle \sum \eta_{(i)}=f_T\]
As in the case with the Dirichlet distribution, the Proposition 21 in \cite{PY} states that sum of the jumps  
$S \sim \Gd(\te,1)$. 
The second statement of the proposition is that $\xi_{(i)}=\eta_{(i)}/S$ are independent of the sum $S$. Finally, sequence of normalized jumps $\xi_{(1)} \ge \xi_{(2)} \ge \dots $ has the Poisson-Dirichlet distribution with parameters $(\al,\te)$.
In what follows Prop. 21 provides convenient way of modeling stochastic evolution of stock prices 'restored' from dynamics of market weights.

\subsection{\pd-market model}
It is natural to employ stick-breaking and size-biased sampling methods 
described in Sections \ref{sec-GDSBP} and \ref{sec-PDD}
for modeling diffusion with stationary Poisson-Dirichlet distribution. 
At first this approach was proposed by Feng and Wang \cite{fengwang}, who also proved reversibility of corresponding infinite-dimensional process. 
Let us recall that the Wright-Fisher diffusion process $Z \equiv Z(t)$ driven by the  SDE
\[
	d Z=\tfrac12\big[\al_1(1-Z)-\al_2 Z\big] d t+\sqrt{Z(1-Z)} dB 
\]
has reversible stationary beta distribution $Z^* \sim \Bd(\al_1,\al_2)$. 

If $X_n(0)$ denotes market weight of the $n$-th largest stock at time $t=0$, then stochastic evolution of market weights can be determined from the stick-breaking  process
\[
	X_1(t)=Z_1(t), \qquad X_n(t)=Z_n(t)\Big(1-\textstyle\sum_{i=1}^{n-1} X_i(t)\Big),
\]
where processes $Z_n \equiv Z_n(t)$ are determined by independent SDEs
\[
	\dif Z_n=\tfrac12\big[(1-\al)(1-Z_n)-(\te+\al n) Z_n\big] \dif t
	+\sqrt{Z_n(1-Z_n)} \dif B_n
\]
with stationary beta distributions, corresponding to the size-biased sampling definition \eqref{TPSB} 
\[Z_n^* \sim \Bd(1-\al,\te+n\al)\]
Initial values of processes $Z_n(0)$ are determined by
\[
	Z_1(0)=X_1(0), \qquad 	
	Z_n(0)=X_n(0)/{(1-\textstyle\sum_{i=1}^{n-1} X_i(0))}
\]
{\em Local} evolution of overall market capitalization $M\equiv M(t)$ can be modelled by diffusion 
\[
	dM=\tfrac12 \big[\te-cM\big]dt+\sqrt{M} \dif B
\]
with stationary gamma distribution $M^* \sim \Gd(\te,c)$, where 
variable $c$ 
is defined by condition $M(0)=\EE M^*$. \\
Correspondingly, local behaviour of stock prices is defined by product of independent processes
\[	
	P_n(t)=\frac{1}{q_n} M(t) \cdot X_n(t),
\]
where $q_n$ denotes number of shares outstanding.

\begin{figure}[H]
\centerline{
\includegraphics[width=16.cm, height=13cm,
trim=1.0cm 6.5cm 1.25cm 6.5cm,clip
]{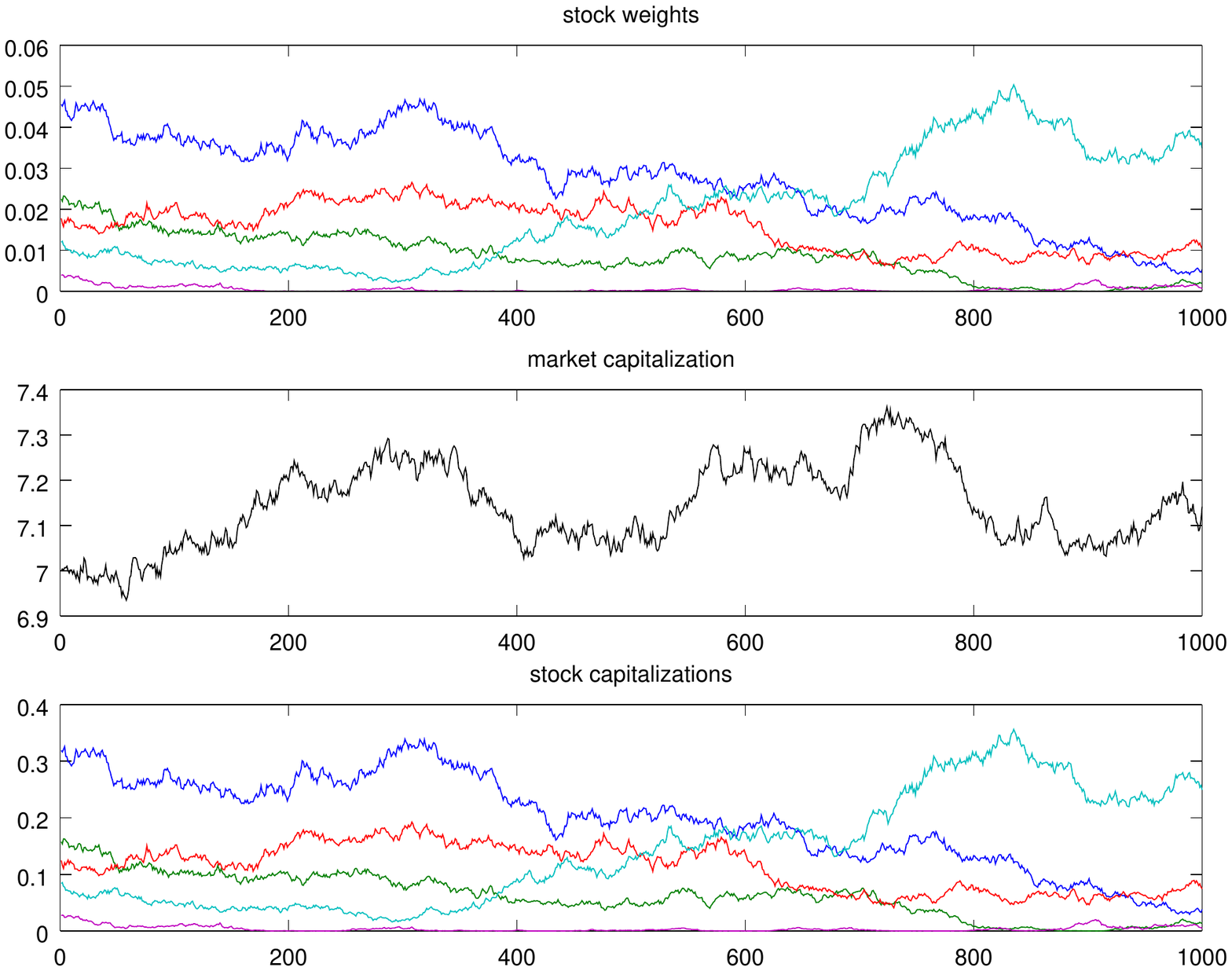}}
\caption{\small Simulation of weights, overall market value and stock capitalizations\\ with stationary
$\mathcal{PD}(0.60, 55)$ distribution}
\end{figure}


\newpage
\subsection{The broken-stick model}\label{sec-brokenstick}
The broken-stick is a simple model illustrating how uniform partition produces inequality patterns. 
MacArthur \cite{macarthur1957relative} proposed this model for explanation of relative species abundances in closed environment. 

Let's assume that stick of unit length represents some finite resource, such as territory, available food, water reservoir, etc., which must be shared between species. The resource is broken at random by throwing uniformly $n-1$ cutting points on this stick and breaking it into $n$ pieces. 
Length of each piece represents share, which is taken by some class of species.
While on average length of each piece will be $1/n$, {\em ranked} lengths of pieces display interesting behavior. 

For instance, if stick is broken just into two pieces, then length of smaller piece is never larger than 50\% and since cut point is uniformly distributed it is easy to see that smaller stick on average represents 25\% of length, while larger one takes 75\%. In general it can be shown that after breaking stick into $n$ pieces expected length of the $k$-th {\em largest} piece is given by
\[
	x_k=\frac 1n\summ jkn \frac 1j
\]
In case of 3 pieces expected proportions ranked in descending order are 
61.1\%, 27.8\% and 11.1\%. It can be checked by straightforward simulation that dropping 4 points at uniform on unit interval produces on average following ranked lengths of 5 subintervals 
\[
	(46\%,26\%,16\%,9\%,4\%)
\]
Obviously, sampled proportions will fluctuate around these expected lengths.
For larger values of $n$ ranked expected proportions start to decay rapidly and it is more convenient to display them on a log-log plot.
\begin{figure}[h]
\centerline{
\includegraphics[width=12cm, height=9cm
]{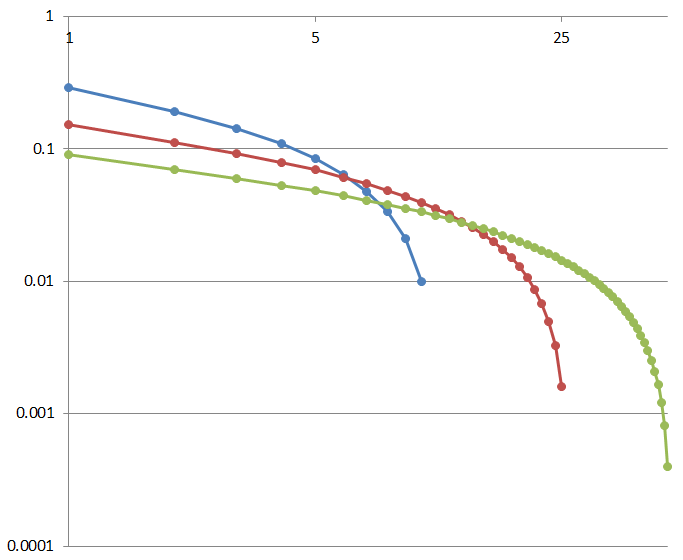}}
\caption{\small Expected proportions for $n=10, 25,50$}
\end{figure}

This example illustrates that asymmetry in ranked proportions  appears with completely uniform distribution of resource.

\newpage
\subsection{Toy model}\label{susec-toy}

Let us imagine that there are only two stocks with capitalizations 3 and 2 in the market with capitalization 5. Tickers or names do not play important role and used only to distinguish the stocks.
Ten ways in which 5 units of money can form a state with these capitalizations is represented by  the ten Young tableaux shown on the left
\begin{table}[H]
\begin{minipage}{.7\linewidth}
\begin{center}
\ytableausetup{smalltableaux}
\ytableaushort{123,4\np}  \quad \ytableaushort{124,3\np} \quad \ytableaushort{134,2\np} 
\quad \ytableaushort{1\np,234}  \\ \vskip 0.2cm
\ytableaushort{12\np,34} \quad \ytableaushort{12,34\np} \quad 
\ytableaushort{13\np,24} \quad \ytableaushort{13,24\np}   \quad 
\ytableaushort{14\np,23} \quad \ytableaushort{14,23\np}
\end{center}
\end{minipage}
$\Bigg \Vert \qquad$
\begin{minipage}{.1\linewidth}
\ydiagram[*(blue!14)]{3,2} 
\end{minipage}\small 
\end{table}
Since these partitions have {\em the same block sizes} it is convenient to use Young {\em diagram}, shown on the right, to denote all partitions with the same {\em shape}.
The 10 partitions above arise by adding a new box:

\begin{minipage}{.45\linewidth}
\ytableausetup{boxsize=.45em}
\begin{itemize}
\item in one way to 4 partitions with shape \raisebox{0.35em}{\hspace{.0em} \ydiagram[*(blue!2)]{3,1}} { }  and\\ 
\item in two ways to 3 partitions with shape \raisebox{0.4em}{\hspace{.0em}\ydiagram[*(blue!3)]{2,2}}
\end{itemize}
\end{minipage}
\hfill
\begin{minipage}{.5\linewidth}
\[\centering
\ytableausetup{boxsize=7pt}
\xymatrix{  
\mu=4,\, \ydiagram[*(blue!1)]{3,1}\ar[dr]		&  		\\
\mu=3,\, \ydiagram[*(blue!1)]{2,2}\ar[r]|2	& \ydiagram[*(blue!15)]{3,2}
}
\]
\end{minipage}
\vskip 0.2cm
For a Young diagram  $\vc c$
combinatorial formula \eqref{numc} 
provides  $\mu(\vc c)$ equal to
\begin{itemize}
	\item number of set partitions with given block sizes, 
	\item which in financial terms is the same as number of ways the market can form a state described by ranked capitalizations.
\end{itemize}
 Exchangeable probability distribution on partitions assigns {\em  the same probability to all partitions with the same shape}. This framework is useful when one is interested in studying distribution of ranked block sizes (capitalizations) regardless of block labels (tickers). 
If $\sfpi n(\vc c)$ denotes probability of a partition with $n$ elements and shape $\vc c$,
then total probability of all partitions with this shape is
\[
	{\sfp n}(\vc c)= \mu(\vc c) \cdot {\sfpi n}(\vc c)
\]
Obviously sum of these probabilities over all shapes (Young diagrams) with $n$ elements must be 1.

In a context of mathematical genetics Kingman \cite{kingman1978random}  considered family of distributions $\{\sfp n\}$ on partitions of $n=1,2,3..$ elements and noticed that random sampling induces natural consistency constraints connecting  distributions  for levels $n-1$ and $n$. He called  distributions $\{\sfp n\}$ satisfying such constraints {\em partition structures}. 
\vskip 0.05cm
\ytableausetup{boxsize=.45em}
Continuing the example, let's consider 10 partitions with the shape \raisebox{0.35em}{\hspace{.0em}\ydiagram[*(blue!3)]{3,2}}. In each of these partitions any of 5 boxes can be removed so remaining partitions will have 4 elements. 
For each partition there are 

\begin{minipage}{.55\linewidth}
\ytableausetup{boxsize=.4em}
\begin{itemize}
\item 2 ways to get to \raisebox{0.35em}{\hspace{.0em}\ydiagram[*(blue!3)]{3,1}}\\
\item 3 ways to obtain \raisebox{0.35em}{\hspace{.0em}\ydiagram[*(blue!3)]{2,2}}{ } 
\end{itemize}
\end{minipage}
\begin{minipage}{.4\linewidth}
\[\centering
\ytableausetup{boxsize=7pt}
\xymatrix{  
\ydiagram[*(blue!2)]{3,1}	& \\ 
\ydiagram[*(blue!2)]{2,2}	& \ydiagram[*(blue!15)]{3,2}\ar[l]|{3/5}\ar[ul]|{2	/5}
}
\]\end{minipage}
\vskip 0.2cm

Uniform deletion of a box on partitions of 5 elements induces probability distribution on 4-partitions. For example
\[\ytableausetup{boxsize=.45em}
	\sfp 4 \Big(\raisebox{0.35em}{\ydiagram{2,2}}\Big)=
	\frac 35 \sfp 5 \Big(\raisebox{0.35em}{\ydiagram{3,2}}\Big)+
	\frac 15 \sfp 5 \Big(\raisebox{0.5em}{\ydiagram{2,2,1}}\Big)
\]
On the other hand this consistency constraint defines {\em forward} conditional probabilities of partition growth, for instance
\[\ytableausetup{boxsize=.45em}
	{\sf p}\Big(\raisebox{0.35em}{\ydiagram{2,2}} \longrightarrow 
				  \raisebox{0.35em}{\ydiagram{3,2}}
			\Big)=
	\frac 35 \sfp 5 \Big(\raisebox{0.35em}{\ydiagram{3,2}}\Big)\Big/
	\sfp 4 \Big(\raisebox{0.35em}{\ydiagram{2,2}}\Big)
\]
\vskip 0.2cm
In other words partition structure is a family of distributions on partitions 
consistent under growth (or Up moves) and recession (or Down moves).
This enables considering market dynamics as a process of {\em combinatorial random walks} on partitions driven by sequences of up or down transitions.

\newpage
\section{Exchangeable partitions} \label{sec-EPS}




\subsection{Descriptions of set partitions}\label{susec-partvec}
Galaxies, stars, companies, people form clusters and sizes of these clusters are rarely uniform. In finance and economics stocks and companies can be considered from the following point of view.
\begin{itemize}
	\item Stock market comprise $k$ stocks with total capitalization $n$ of units of money. If $n_i$ denotes value of $i$-th largest company (by capitalization), then market weights are given by $x_i=n_i/n$ and vector $\vc x=(x_1, x_2, ..)$ represents {\em capital distribution curve}. 

	New unit of money can join any of the stocks thus increasing capitalization of particular stock to $n_i+1$ and total capitalization to $n+1$, or unit of money can leave a stock decreasing corresponding stock and market capitalizations by 1. Also there is a possibility that a new stock will be issued during the IPO, which leads to increase of number of clusters to $k+1$.
	\item In the same way companies assets/values may experience increase or decrease. Also there is a possibility that a new company enters the market. 
	\item In mutual funds industry, money coming to the market join existing funds proportionally to their size, but there is always opportunity that new fund emerges.
\end{itemize}

Process of clustering can represented by partitions of a set. For instance, the set of three letters 'a', 'b' and 'c' can be partitioned as shown below in the left column, with corresponding Young diagrams in the right column representing {\em partition classes} :
\begin{center} 
\begin{tabular}{c | c}
\ytableausetup{boxsize=0.9em}
$\{a,b,c\}$ &	\ydiagram[*(blue!4)]{3}  \Bs \\   \hline  \Ts
$\{a,b\}, \{c\}$ & 	\multirow{3}{*}{\ydiagram[*(blue!4)]{2,1}} \Bs \\
$\{a,c\}, \{b\}$ &	\\
$\{b,c\}, \{a\}$ &	 \\\hline \Ts
\multirow{2}{*}{$\{a\}, \{b\}, \{c\}$} &	\ydiagram[*(blue!4)]{1,1,1} 
\end{tabular}
\end{center}

 If in set partition, represented by clusters/blocks,  cluster labels are not important and order of items inside of each cluster is irrelevant, then such partition called {\em exchangeable}. Such partitions have the same {\em shape} and completely described by vector of their block sizes. Partitions with the same shape belong to the same {\em exchangeable  class} (or {\em partition class}). For the example with three companies, the set $\{a,b,c\}$ has 5 partitions and 3 exchangeable classes, represented by Young diagrams in the right column.


Every exchangeable partition of $n$ elements into $k $ clusters (blocks) can be described in two ways.
\begin{enumerate}
	\item For the first order description, since labeling of clusters is not important,  it is convenient to consider cluster sizes arranged in descending order 
		\[
			n_{1} \ge n_{2}\ge \dots \ge n_{k},
		\]
		where $n_i$ denotes size of $i$-th largest cluster, hence $n=n_1+\dots+n_k$\\
		In population biology terminology it is called a {\em frequency vector:}
		\[ \vc n=[n_1, n_2, \dots n_k]\]
		Obviously Young diagrams correspond to this description. 
	\item For the second description, let $c_1$ denote number of clusters of size one,  
	$c_2$ denote number of clusters of size two, etc. If $c_j$ denote number of clusters with $j$ items then total number of items is
	\[
		n=c_1+2 \cdot c_2+\dots +m \cdot c_m	
	\]
	and number of clusters is given by $k=c_1+\dots +c_m$
	where $m$ is the size of the largest cluster. 
	To distinguish {\em partition vector} $\vc c$  from frequency vector  $\{\}$-notation is used 
	\[
		\vc c=\{c_1, c_2, \dots \}
	\]	
Let $\vc c \prt(n,k)$ denote that vector $\vc c$ describes a partition of $n$ elements into $k$ clusters.
\end{enumerate}

\newpage
\subsection{Size of \exc class}\label{combf}

Every \exc class contains set partitions, represented by the same partition vector $\vc c=\{c_1, c_2, c_3,...\}$.
Number of partitions in a class, described by vector $\vc c$, is given by 
\eqn\label{numc}
	{\mu}(\vc c)=\frac{n!}{\prod c_j ! (j!)^{c_j}}
\nqe
Indeed, by multinomial formula number of partitions with $c_1$ one-element subsets, $c_2$ two-element subsets, etc. is 
\[
	\frac{n!}{\underbrace{1!\dots1!}_{c_1 \text{ times}}
		\underbrace{2!\dots2!}_{c_2 \text{ times}} \dots
	}
\]
which must be divided by $\prod c_j! $ since permutations of blocks of the same size play no role.\\
For instance, the 4 element set $\{a,b,c,d\}$ has 15 partitions and 5 exchangeable classes:

\begin{center} 
\begin{tabular}{c | c | c | c | c}
\ytableausetup{boxsize=0.9em}
$\{a,b,c,d\}$ &	\ydiagram[*(blue!4)]{4}  
& $\vc n=[4]$ 
& $\vc c=\{0,0,0,1\}$ 
& $\displaystyle\mu(\vc c)=\frac{4!}{(4!)^1}=1$
\Bs \Bs\\  
\hline  
\Ts
$\{a,b,c\}, \{d\}$ 
& \multirow{4}{*}{\ydiagram[*(blue!4)]{3,1}}  
& \multirow{4}{*}{$\vc n=[3,1]$ }
& \multirow{4}{*}{$\vc c=\{1,0,1,0\}$} 
& \multirow{4}{*}{$\displaystyle\mu(\vc c)=\frac{4!}{(1!)^1(3!)^1}=4$} \\ 
$\{a,b,d\}, \{c\}$ &&&&	\\
$\{a,c,d\}, \{b\}$ &&&&	\\
$\{b,c,d\}, \{a\}$ &&&& \\
\hline \Ts
$\{a,b\}, \{c,d\}$ 
& \multirow{3}{*}{\ydiagram[*(blue!4)]{2,2}} 
& \multirow{3}{*}{$\vc n=[2,2]$} 
& \multirow{3}{*}{$\vc c=\{0,2,0,0\}$} 
& \multirow{3}{*}{$\displaystyle\mu(\vc c)=\frac{4!}{2! (2!)^2}=3$}
\Bs \\
$\{a,c\}, \{b,d\}$ &&&&	\\
$\{a,d\}, \{b,c\}$ &&&&	 \\
\hline \Ts
$\{a,b\}, \{c\},\{d\}$ 
& \multirow{6}{*}{\ydiagram[*(blue!4)]{2,1,1}} 
& \multirow{6}{*}{$\vc n=[2,1,1]$} 
& \multirow{6}{*}{$\vc c=\{2,1,0,0\}$} 
& \multirow{6}{*}{$\displaystyle\mu(\vc c)=\frac{4!}{2! (1!)^2 (2!)^1}=6$}
\Bs \\
$\{a,c\}, \{b\},\{d\}$ &&&&\\
$\{a,d\}, \{b\},\{c\}$ &&&&\\
$\{b,c\}, \{a\},\{d\}$ &&&&\\
$\{b,d\}, \{a\},\{c\}$ &&&&\\
$\{c,d\}, \{a\},\{b\}$ &&&&\\
\hline \Ts
\multirow{3}{*}
{$\{a\}, \{b\}, \{c\}, \{d\}$} &	\ydiagram[*(blue!4)]{1,1,1,1} 
& \multirow{3}{*}{$\vc n=[1,1,1,1]$} 
& \multirow{3}{*}{$\vc c=\{4,0,0,0\}$} 
& \multirow{3}{*}{$\displaystyle\mu(\vc c)=\frac{4!}{4! (1!)^4}=1$}
\end{tabular}
\end{center}
Interestingly that partition $\vc n=[2,1,1]$ can be realized in 6 ways and uniform partition $\vc n=[2,2]$ only in 3 ways. 

\subsection{Partition structure} 

If all partitions from the class with partition vector $\vc c$ are considered to be equivalent  then they should have the same probability. 

If $\sfpi n (\vc c)$ denotes probability of an element from the partition class $\vc c$ with $\mu(\vc c)$ elements then probability of that \exc class is 
\[
	\sfp n(\vc c)= \mu(\vc c) \cdot {\sfpi n}(\vc c)
\]
Obviously these probabilities should satisfy
\eqn\label{exch-prob}
	\sum_{\vc c \prt (n) } {\sfp n}(\vc c)=1
\nqe
Here $\vc c \prt (n)$ denotes that summation runs over all classes of partitions of $n$ elements. 

\prg{Partition structure.}
In general, it is not enough to assign probability measures over partition classes for all values of $n$. 
Kingman \cite{kingman1978random} noticed that besides \eqref{exch-prob} there are consistency conditions linking exchangeable probability measures ${\sfp {n-1}}$ and $\sfp n$ and called such consistent sequences of $\{\sfp i\}$  {\em partition structures}.

\newpage
\subsection{Ewens-Pitman Sampling Formulae}\label{EPSF}
A finite dimensional counterpart of one-parameter stick-breaking model \eqref{OPSB} has been proposed by Ewens in the context of population biology.
Given partition vector $\vc c$ Ewen's Sampling Formula assigns probability as
\eqn\label{ESF}
	p(\vc c)=\frac{\te^k}{\rfl \te n} \frac{n!}{\prod c_j ! j^{c_j}}
\nqe
Pitman studied size-biased representation of the two-parameter model \eqref{TPSB} and obtained corresponding extension of Ewen's sampling formula in \cite{pitman1995exchangeable}. The two-parameter Pitman's Sampling Formula (PSF) gives probability for partition class $\vc c$
\eqn\label{PSF}
	p(\vc c)=\frac{\rfl \te {k,\al}}{\rfl \te n} \frac{n!}{\prod c_j!}
	\prod \bigg(\frac{\rfl {1-\al}{i-1}}{i!} \bigg)^{c_i}
\nqe
where $\rfl \te {k,\al}=\te(\te+\al)\cdots(\te+\al(k-1))$, which shows that for $\al=0$ the formula converges to \eqref{ESF}. 

Kerov\cite{kerov2006coherent} proposed that formula \eqref{PSF} can be obtained via  model of random allocation with conditionally independent variates.

\subsection{Chinese Restaurant Process}\label{susec-CRP}
Chinese Restaurant Process provides probabilistic dynamics of partitions, ensuring that probabilities remain exchangeable.
Zabell \cite{zabell2005symmetry} explains the metaphor  as 
{\em ''on any given evening in Berkley a large number of people go to some Chinese restaurant in the downtown area. As each person arrives, he looks in the window of each restaurant to decide whether or not to go inside. His {\cldb chances of going in  increases with the number of people already seen inside}...
But { \cldb there's some probability that he goes to an empty restaurant..}''} 

More formally, it is assumed that there are infinite number of tables (restaurants)
and first customer always sits at first unoccupied table, say table 1. Customer $n+1$ observes occupied $k$ tables and
	\begin{itemize}
		\item joins table with $n_i$ people with probability
		 	 \[ p_i=\frac{n_i -\clb \al}{n+\te} \]
		\item joins new, unoccupied table with probability 
		{\[ p^*=\frac{{\clr \te} + \al k }{n+ \te} \]}
	\end{itemize}
It is important that this process provides {\em exchangeable} probability on partitions.


For instance partition $\vc n=[2,1,1]$ can migrate to following states

\begin{minipage}[r]{0.3\textwidth}
\begin{tikzpicture}
	\node (c0) at (1,8) [circle,minimum size=0cm,draw, color=white] {};
	\node (c1) at (5,7) [circle,minimum size=1.25cm,draw] {$\frac {2-\clb{\al}} {\te+4}$}; 
	\node (c2) at (5,5) [circle,minimum size=1.25cm,draw] {$\frac {1-\clb{\al}} {\te+4}$}; 
	\node (c3) at (5,3) [circle,minimum size=1.25cm,draw] {$\frac {1-\clb{\al}} {\te+4}$}; 
	\node (c4) at (5,1) [circle,minimum size=1.25cm,draw] {$\frac {{\clr\te}+3\clb\al} {\te+4}$}; 
	\shade[shading=ball, ball color=blue!50] (c1.95)  circle (.11);
	\shade[shading=ball, ball color=blue!50] (c1.225)  circle (.11);
	\shade[shading=ball, ball color=blue!50] (c2.225)  circle (.11);
	\shade[shading=ball, ball color=blue!50] (c3.225)  circle (.11);

\end{tikzpicture}
 \end{minipage}
 \begin{minipage}[]{0.4\textwidth}
\[\small
{
\xymatrix{  
	& \ydiagram[*(blue!40)\bullet]{2+1} *[*(blue!4)]{3,1,1} \\
\ydiagram[*(blue!4)]{2,1,1}\ar[ur] \ar[r]|2 \ar[dr] & \ydiagram[*(blue!40)\bullet]{0,1+1} *[*(blue!4)]{2,2,1} \\
	& \ydiagram[*(red!40)\bullet]{0,0,0,1} *[*(blue!4)]{2,1,1,1} \\
}
}
\]
 \end{minipage}

\newpage
\subsection{Infinite dimensional diffusions as random walks on partitions}\label{sec-diffus}

Diffusion processes in the ordered infinite simplex were developed by Petrov, Olshanski, Borodin (\cite{borodin2009infinite},\cite{petrov2009two}, also \cite{fulman2005stein}), by Feng et al. (\cite{fengwang}, \cite{feng2010some} and \cite{feng2010poisson}) and Ruggiero and Walker (\cite{ruggiero2009countable}). Independently, Markov chain induced by down- up- transitions was studied by Costantini, Garibaldi et. al.  (\cite{garibaldi2003exact}, \cite{garibaldi2007two}).

The idea of approximating a two-parameter diffusion process is relatively simple. Let's fix some large $n$ and starting with some partition let's consider Down and Up jumps, where
\begin{itemize}
\item down move randomly deletes one cell in Young diagram
\item up move acts according to the Chinese Restaurant Process
\end{itemize}

For instance, figure below shows possible transitions between partitions of 4 and corresponding DU-chain:

\begin{figure}[H]
\centering
\begin{subfigure}{.3\textwidth}
  \centering
  \[
\ytableausetup{boxsize=6pt}
{%
\xymatrixcolsep{12mm}
\xymatrixrowsep{1.4pc}
\xymatrix{  
 					& \ydiagram[*(blue!10)]{4} 		\ar[dl]	\\
\ydiagram{3} 	& \ydiagram[*(blue!10)]{3,1} 	\ar[l]  \ar[dl]|3	\\
\ydiagram{2,1}  	& \ydiagram[*(blue!10)]{2,2}	\ar[l]			\\
\ydiagram{1,1,1}& \ydiagram[*(blue!10)]{2,1,1} \ar[l]|2 \ar[ul]|2		\\
					& \ydiagram[*(blue!10)]{1,1,1,1} \ar[ul]
}
}
\]
  \caption{{\sc Down}-chain}
  \label{fig:ub1}
\end{subfigure}%
\begin{subfigure}{.3\textwidth}
  \centering
\[
\ytableausetup{boxsize=6pt}
{%
\xymatrixcolsep{12mm}
\xymatrixrowsep{1.4pc}
\xymatrix{  
 									& \ydiagram[*(blue!10)]{4} 	\\
\ydiagram{3}\ar[r]  \ar[ur] 	& \ydiagram[*(blue!10)]{3,1} 		\\
\ydiagram{2,1}\ar[r]\ar[dr]\ar[ur]		& \ydiagram[*(blue!10)]{2,2}				\\
\ydiagram{1,1,1} \ar[r]|3 \ar[dr] 	& \ydiagram[*(blue!10)]{2,1,1} 		\\
									& \ydiagram[*(blue!10)]{1,1,1,1} 
}
}
\]
  \caption{{\sc Up}-chain}
  \label{fig:ub2}
\end{subfigure}
\begin{subfigure}{.3\textwidth}
  \centering
\[
\ytableausetup{boxsize=6pt}
{%
\xymatrixcolsep{12mm}
\xymatrixrowsep{1.4pc}
\xymatrix{  
 \ydiagram[*(blue!10)]{4} 		\ar@{~}[d] & 	\\
 \ydiagram[*(blue!10)]{3,1} 	 \ar@{~}[dd] & \\
& \ydiagram[*(blue!10)]{2,2}	\ar@{~}[dl] \ar@{~}[ul]		\\
 \ydiagram[*(blue!10)]{2,1,1} 	\ar@{~}[d] &	\\
\ydiagram[*(blue!10)]{1,1,1,1} &
}
}
\]
  \caption{DU-chain}
  \label{fig:ub2}
\end{subfigure}
\caption{\small Example of Down-Up transitions }\label{}
\label{fig:test}
\end{figure}
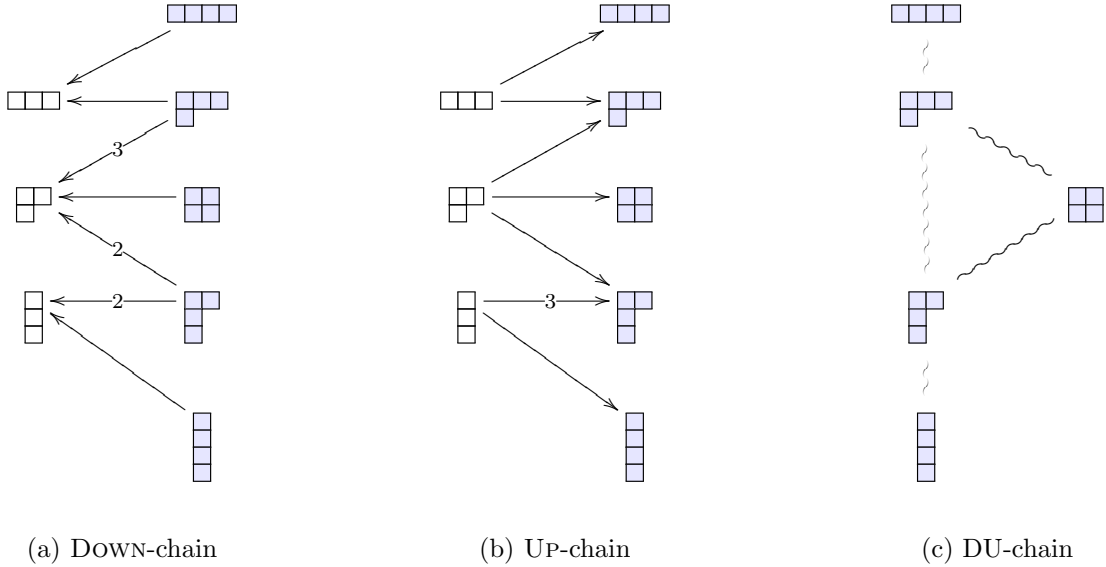
D- and U- operators preserve probability structures given by Ewens-Pitman Sampling formulae, hence obtained Markov chain also preserves this distribution.
Figure below illustrates approximation of diffusion process in Kingman simplex with parameters $\al=0.3, \te=5$
\begin{figure}[H]
\centering
  \includegraphics[width=14.5cm, height=7.25cm,
	trim=1.5cm 7cm 2cm 6.7cm,clip]{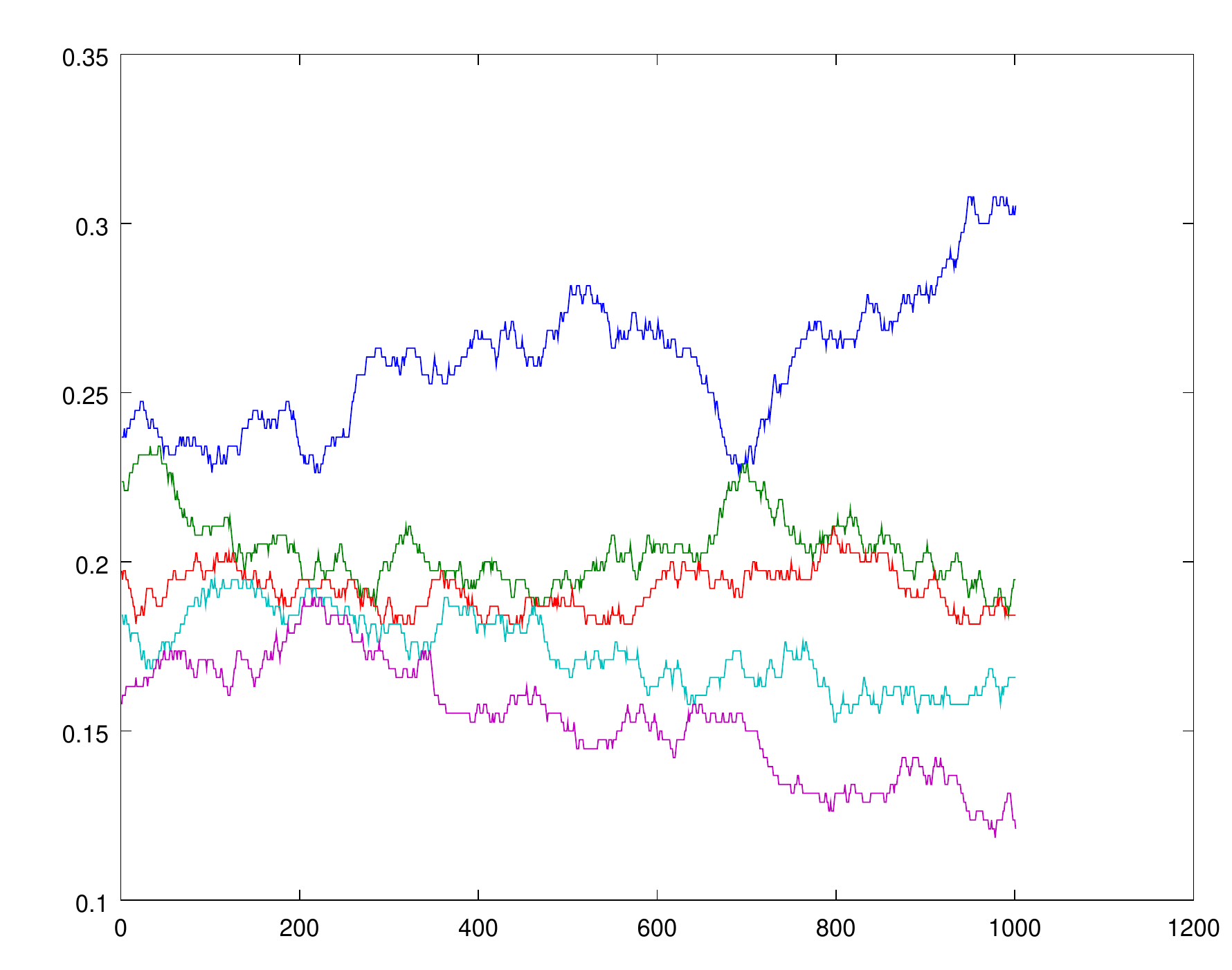}
\caption{\small Diffusion sample paths on partitions, top five $x_i(t)$}\label{diffus}
\end{figure}

\newpage
\section{Dirichlet distribution and size-biased sampling}\label{sec-GDSBP}

\subsection{Dirichlet distribution}
The broken stick model is a particular case of more general Dirichlet distribution, which is a probability measure over vectors of proportions. Its density function is parametrized by vector 
$\vcc \al m$ 
\[
	p(\vc x)=\frac{\GG(\sum \al_i)}{\GG(\al_1)\cdots\GG(\al_m)} 
	x_1^{\al_1-1} \cdots x_m^{\al_m-1}
\]
The distribution is defined for vectors of proportions from the simplex
\[
	\Delta_m=\big\{ \vc x \in \RR^m | \textstyle\summ i1m x_i=1, \; x_i \ge0 \big\}
\]
It is convenient to denote that vector $\vc x \in \Delta_m$ has $m$-dimensional Dirichlet distribution described by vector of parameters $\vc \al$ as
\[
	\vc x \sim \Dir_m(\vc\al)
\]
An alternative and equivalent definition of the Dirichlet distribution, which makes proofs of its properties almost immediate is given by normalization of vector of gamma variables. For $m$ independent gamma distributed variables $y_i \sim \mathcal G (\al_i)$ vector of proportions defined by
\[
	\vc x=\bigg( \frac {y_1}{\sum y_i}, \cdots , \frac {y_m}{\sum y_i} \bigg)
\]
has Dirichlet distribution  $\vc x \sim \Dir_m(\vc\al)$. 

Gamma distribution possess convolution property \[
	\text{for } y_1 \sim \mathcal G (\al_1), y_2 \sim \mathcal G (\al_2) 
	\qquad y_1+y_2\sim \mathcal G (\al_1+\al_2)
\]
This property together with 
Lukacs characterization, saying that it is unique distribution which possesses independence of $y_1/y_2$ from $y_1+ y_2$,
 simplifies proofs of the properties below.
\prg{Properties.} 
Let $\te=\summ i1m \al_i$ denote sum of parameters. If $y_o=\summ j1m y_j$ denotes sum of independent gamma variables  $y_i \sim \mathcal G (\al_i)$, then this sum has gamma distribution $y_o \sim \mathcal G(\te)$ by convolution property.

If component $y_i$ is separated and others are lumped together, then vector $(y_i, \sum_{j\neq i} y_j)$ has independent components with gamma distributions with parameters $\al_i$ and $\te-\al_i$ correspondingly. From normalization by $y_o$ it follows that
marginally $x_i=y_i/y_o$ have beta distributions
\[
	x_i \sim \BB(\al_i, \te-\al_i)
\]
Let $\vc x_{[-i]}$ denote that $i$-th component is removed from the vector $\vc x$
\[
	\vc x_{[-i]}=(x_1,... x_{i-1}, x_{i+1},... x_m)
\]
Since in a vector $\vc x \in \Delta_m$ sum of components is 1, in a vector $\vc x_{[-i]}$ sum of components becomes $1-x_i$ and therefore normalized vector 
$\vc x_{[-i]}/(1-x_i)$ belongs to $\Delta_{m-1}$ simplex.\\
The Dirichlet distribution possesses an important property of {\em neutrality} that is independence of $x_i$ and $\vc x_{[-i]}/(1-x_i)$.
Moreover, if original vector $\vc x \sim \Dir_m(\vc \al)$, then 
\[
	\frac{\vc x_{[-i]}}{1-x_i} \sim \Dir_{m-1}(\vc \al_{[-i]})
\] 
This property is trivial for $m=2$. For arbitrary dimension, let's consider vector with $m$ independent gamma variates $y_i$ and corresponding Dirichlet distributed vector with components 
$x_j=y_j/y_o$. Removing $i$-th component from both vectors and normalization of the second one yields for $j\neq i$
\[x_j \cdot \frac1{1-x_i}=x_j \cdot \frac1{1-y_i/y_o}=\frac {y_j}{y_o}\frac{y_o}{y_o-y_i}=\frac{y_j}{ \sum_{ k\neq i} y_k}\]
Independence (neutrality) follows from Lucacs property.

\prg{Beta stick-breaking.} 

These properties suggest a {\em stick-breaking} method of sampling from the Dirichlet distribution. Let's imagine that a stick of unit length is broken by the following step-by-step procedure.
By marginal property the first component $x_1=z_1$ can be modeled as 
\[z_1 \sim \BB(\al_1, \te-\al_1)\]
By neutrality property the first piece  can be broken off. 
The remaining components have $\Dir_{m-1}(\vc \al_{[-i]})$ distribution.
The procedure is repeated for remaining part of stick with length $1-x_1$ with 
\[	{\clb z_2} \sim \BB(\al_2, \te-\al_1-\al_2) \]

\begin{center}
\begin{tikzpicture}[scale=1]
\draw 	(0.0,0) rectangle node[yshift=-0.03cm]{\footnotesize $x_1$} (1.5,0.45);
\draw 	(1.5,0) rectangle node[yshift=-0.01cm]{\footnotesize $x_2=z_2(1-z_1)$} (4.0,0.45);
\draw 	(4.0,0) rectangle node{} (8.0,0.45);

\draw [decorate,decoration={brace,amplitude=5pt,},xshift=0pt,yshift=0pt]
(1.5,0.45) -- (4,0.45) node [black,midway,below,xshift=0.0cm, yshift=0.58cm] 
{\footnotesize $\clb z_2$};
\draw [decorate,decoration={brace,amplitude=5pt,},xshift=0pt,yshift=0pt]
(4,0.45) -- (8,0.45) node [black,midway,below,xshift=0.1cm, yshift=0.62cm] 
{\footnotesize $1-z_2$};
\draw [decorate,decoration={brace,amplitude=8pt,mirror,raise=1.4pt},xshift=0pt,yshift=0pt]
(1.5,0.0) -- (8,0.0) node [black,midway,below,yshift=-0.3cm] 
{\footnotesize $ 1-z_1$};
\end{tikzpicture}
\end{center}
producing new piece $x_2=z_2(1-z_1)$ and residual with length $(1-z_2)(1-z_1)=1-x_1-x_2 \;$  etc. \\
In general by stage $k$
\begin{align*}
	z_k &\sim \BB(\al_i, \te-\textstyle\summ j1k \al_j) \\
	x_k &=z_k \prod_{j=1}^{k-1} (1-y_j) = z_k \bigg(1-\sum_{j=1}^{k-1} x_j \bigg)
\end{align*}
For $k = m$ the last component is simply a remainder.
In the case of symmetric Dirichlet distribution with $\al_i=\al$ breaking rule simplifies to 
\[
	z_k \sim \BB(\al, \te-k \al) \]\[
	x_k =z_k  (1-x_1-\dots -x_{k-1})
\]

\subsection{Size-biased sampling} \label{sec-SBP}
Stick-breaking method from the previous paragraph samples proportions from the Dirichlet distribution component by component.
In many applications as well as in the development of the Poisson-Dirichlet distribution it  is more important to study proportions ranked in descending order 
\[ x_{(1)} \ge x_{(2)} \ge \dots \]
so it would be more convenient to have a procedure which gives as output order statistics from distribution. In general this is not easy, however it is possible to devise a simulation strategy which provides samples in proportions of the appearance in the real distribution -   which is given by {\em size-biased sampling}.

Let's suppose that given a Dirichlet distributed vector
$\vc x \sim \Dir_m(\vc \al) $ one of its components is chosen at random, such that $x_j$ is the probability of choosing $j$-th component. Alternatively it can be visualized as dropping a point at uniform over stick of length one, which is divided by proportions $x_1,x_2,..$ and choosing proportion/piece at which the point falls on.
Value of this proportion is called {\em size-biased sample} and obviously chances to choose largest piece are highest, etc. Once proportion is chosen, it is set apart and the procedure is repeated with normalized residual. The outcome is the {\em size-biased permutation} of vector $\vc x$ where components are randomly interchanged, with bias towards ordered case.

Density function of the first size-biased pick in the permuted vector can be found by the following argument.
Proportion $\wt x$ may be picked with probability $\wt x$ as a first component of a vector $(\wt x, x_2, ..., x_m)$ or as a second component of a vector $(x_1, \wt x, x_3, ..., x_m)$, etc. In each of these particular cases finding unconditional probability amounts to marginalization over Dirichlet density, which yields beta densities $B_{\al_i,\te-\al_i}(\wt  x)$ and therefore total probability is
\[
	p(\wt x)=
	\summ i1m \wt x \cdot B_{\al_i,\te-\al_i} (\wt x)
\] 
In case of symmetric Dirichlet distribution density of the size-biased pick simplifies to
\[
	p(\wt x)=\wt x m \frac{\al\GG(\al m)}{\al\GG(\al)\GG(\te-\al)} \wt x^{\al-1}(1-\wt x)^{\te-\al-1} 
	=\frac{\GG(\te+1)}{\GG(\al+1)\GG(\te-\al)} \wt x^{\al}(1-\wt x)^{\te-\al-1} 
\]
In other words, first SBP $\wt x_1=\wt y_1$ has beta distribution with shifted parameter
\[
	\wt x_1 \sim \BB(1+\al, \te-\al)
\]
After breaking off the first SBP and applying the procedure over and over again it can be shown that piece to be broken off from the remaining part of the stick has beta distribution
\eqn\label{SBSB}
	\wt z_k \sim \BB(1+\al, \te-\al k) 
\nqe
with corresponding proportions
\[
	\wt x_k= \wt z_k (1-\wt x_1 - \dots -\wt x_{k-1})
\]
Since $\te=\al m$ simulation terminates at stage $k=m-1$ with $\wt x_m$ is length of the remainder.

Samples obtained this way will have tendencies for larger proportions appearing first, followed by smaller proportions. Obviously, after ranking of proportions both methods (standard and size-biased ones) produce identically distributed sequences, since SBP only randomly permutes components of symmetric Dirichlet vector. 

At first it is not clear what is the purpose of the SBP, since components of ordered Dirichlet distributed vector can be sampled by standard procedures and then ranked. However, as it will be shown below, size-biased simulation allows sampling from directly not accessible cases.

\section{The Poisson-Dirichlet distribution}\label{sec-PDD}

\subsection{One-parameter family}

For symmetric $m$-dimensional Dirichlet distribution $\Dir_m(\al)$ let's consider limiting case, where dimensionality $m\to\infty$ such that $\te=\al m$, which means that while dimension goes to infinity the total charge $\te$ remains the same and individual parameters $\al=\frac\te m\to0$. In this case direct application of the standard stick-breaking method is impossible since we will have to sample from $\BB(\eps,\te-\eps)$ where $\eps$ is infinitesimally small. However, for any $m$, it is possible to consider size-biased stick-breaking with
\[
	\wt z_k \sim \BB(1+\te/ m,\te-k \cdot \te / m)
\]
which gives for $m\to\infty$ 
\eqn\label{OPSB}
	\wt z_k \sim \BB(1,\te)
\nqe
\[ 
	\wt x_k= \wt y_k (1-\wt x_1 - \dots -\wt x_{k-1})
\]
Ranked values of size-biased sequence $\{\wt x_k\}$ have one-parameter Poisson-Dirichlet distribution $\mathcal{PD}(\te)$
\[
	\wt x_{(1)} \ge \wt x_{(2)} \ge \dots
\]
It is important to not that in contrast to \eqref{SBSB} sequences of $\wt z_k$ and $\wt x_k$ are infinite, since ultimately they correspond to the infinite dimensional Dirichlet distribution.

An alternative way of sampling consists in consideration of jumps of gamma subordinator in time interval $[0,\te]$ and normalization of ranked jumps in this interval. 

\subsection{Two-parameter family}\label{sec-twop-sbr}

Engen\cite{engen1978} noticed that valid size-biased stick-breaking model \eqref{SBSB} holds for negative values of parameter in the range $\al\in(-1,0]$, which after relabeling $ (-\al) \longmapsto \al$ leads to the following sampling method
\eqn\label{TPSB}
	\wt z_k \sim \BB(1-\al, \te+\al k) 
\nqe
where as before corresponding partitions of unit interval are given by
\[
	\wt x_k= \wt y_k (1-\wt x_1 - \dots -\wt x_{k-1})
\]
The {\em \tpPD distribution} is defined as distribution of ranked values of sequence $\{\wt x_k\}$ 
\[
	\wt x_{(1)} \ge \wt x_{(2)} \ge \dots
\]
Obviously in the setting of \eqref{TPSB} range of parameter $\al$ is $-\te<\al<1$. 
For values $\al<0$ such that $\te=m\al$ sequence $\wt x_k$ eventually stops and corresponds to the Dirichlet distribution .
When $0\le\al<1$ as well as in one-parameter case sequence $\wt x_k$ does not stop and thus for this range of parameter $\al$ this model is infinite-dimensional.

\newpage
\section{Market capitalizations and $\PDat$}\label{sec-examples}

Plots in this section illustrate results of fit of capital distribution curves for several international stock exchanges by averages of the Poisson-Dirichlet law. Blue circles correspond to ranked relative capitalizations 
and red lines represent averages of several samples from the \tpPD distribution. Optimal fitting is by least-squares method. 

The first example shows that the \PD distribution provides fit for ranked normalized capitalizations of world stock exchanges. Data source is {\tt \cldb http://www.world-exchanges.org/statistics/monthly-reports}, data are as of end of November, 2014.

\begin{figure}[H]
\centerline{
\includegraphics[width=12.5cm, height=7cm,
trim=1.5cm 7cm 2cm 6.7cm,clip
]{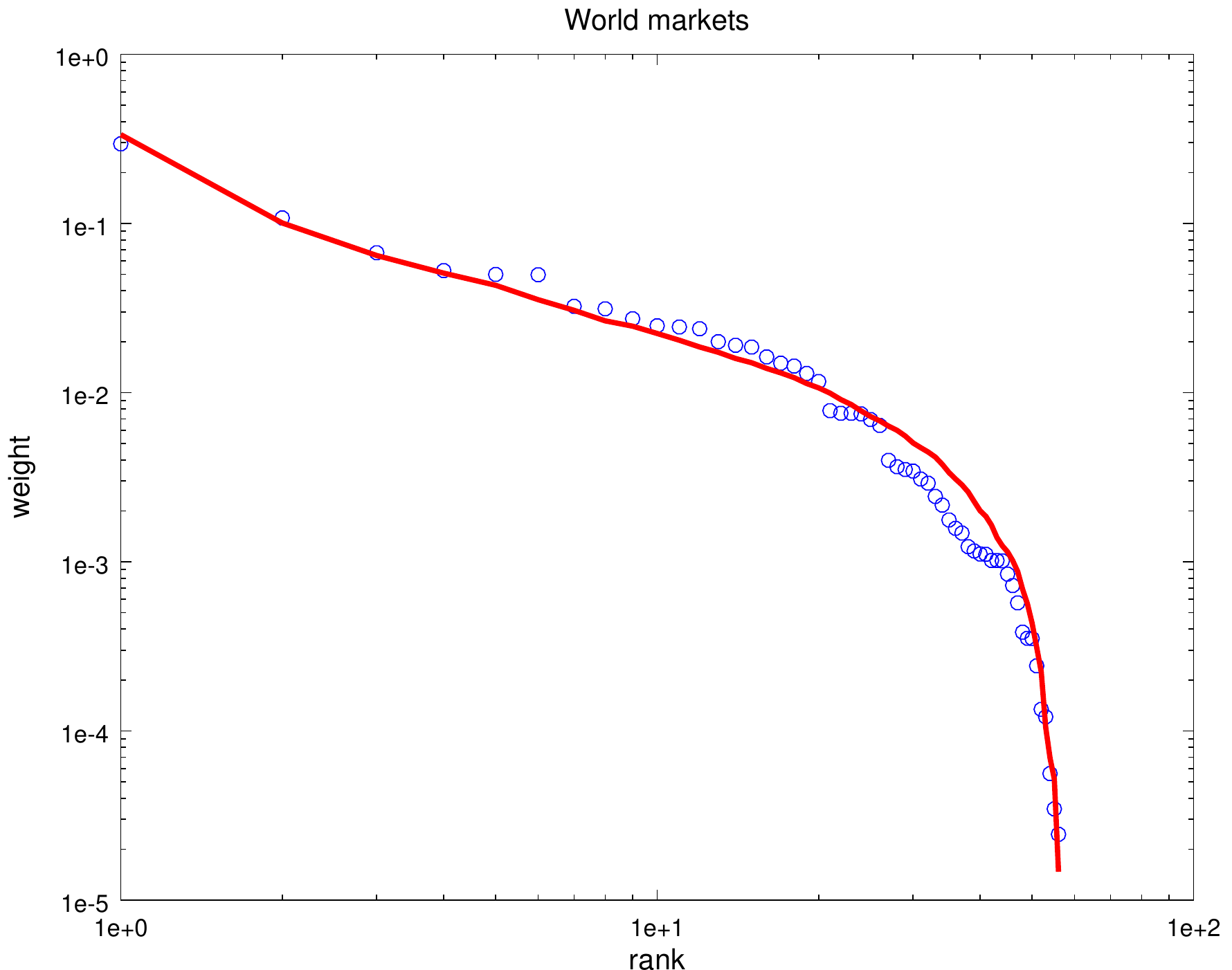}}
\caption{\small NYSE, NASDAQ, Japan, EURONEXT, Hong Kong,... $(\al=0.44,\te=18)$}
\end{figure}

Figures below illustrate results of modeling of  capital distribution curve in major stock exchanges. 
Data source is {\tt \cldb http://www.google.com/finance\#stockscreener}, data are as of  December 9, 2014. Possible explanation of these results is provided by the hypothesis that the two-parameter model approximates underlying partition structure in corresponding markets. This suggests that stochastic evolution of ranked market weights can be modeled by the two-parameter diffusion process or its modification.

\begin{figure}[H]
\centerline{
\includegraphics[width=12.5cm, height=7cm,
trim=1.5cm 7cm 2cm 6.7cm,clip
]{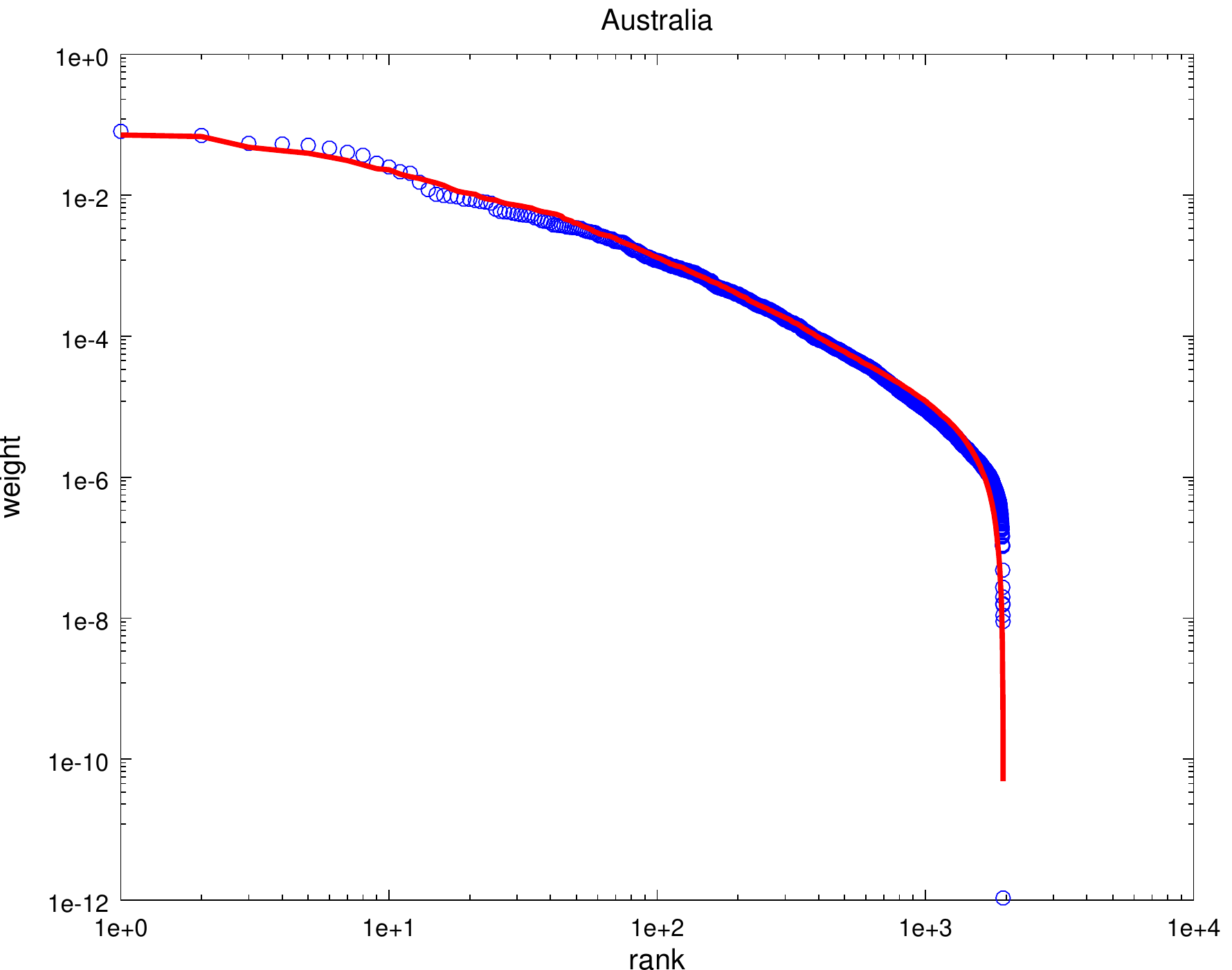}}
\caption{\small Australia $(\al=0.45,\te=18)$}
\end{figure}

\begin{figure}[H]
\centerline{
\includegraphics[width=12.5cm, height=6.75cm,
trim=1.5cm 7cm 2cm 6.7cm,clip
]{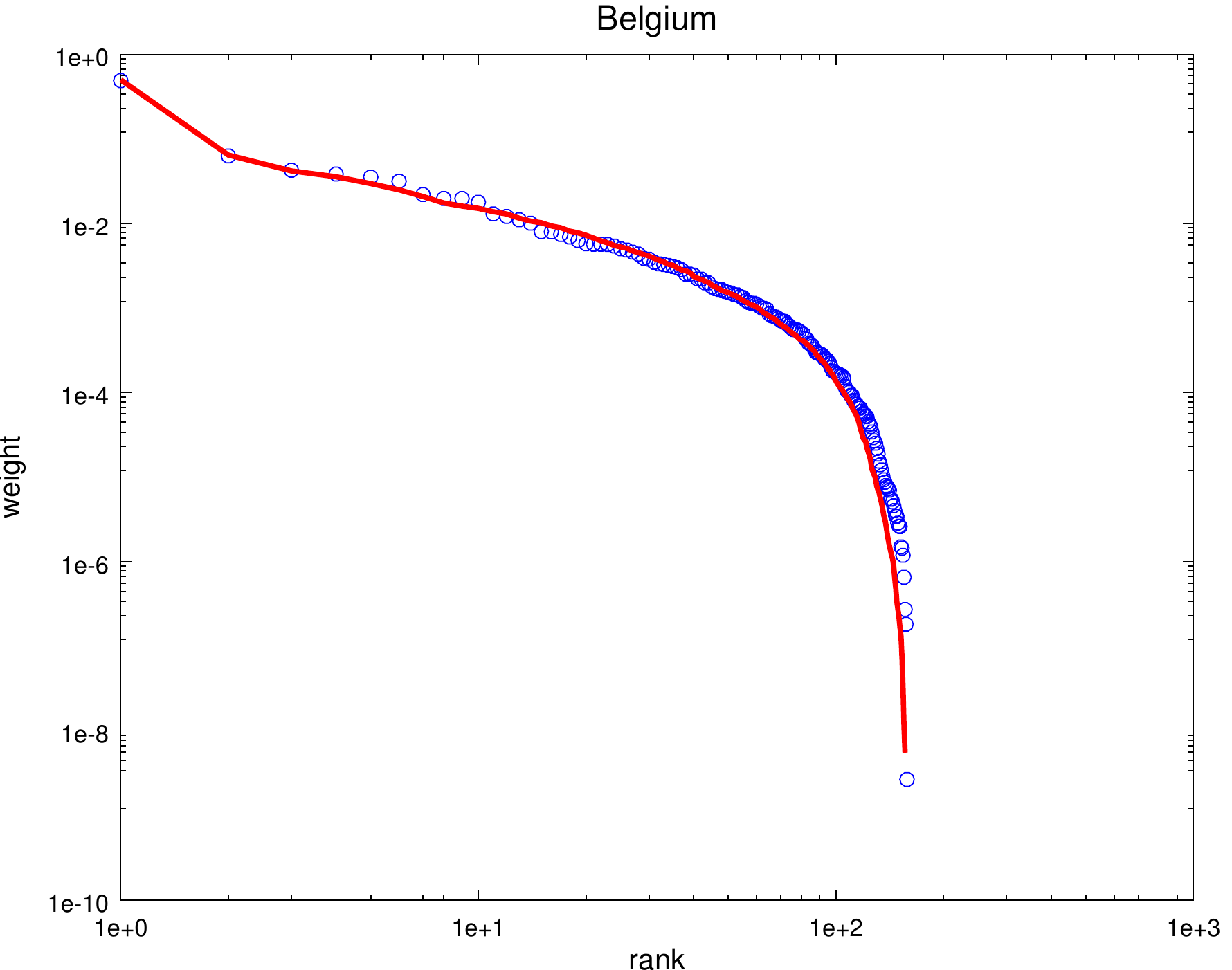}}
\caption{\small Belgium $(\al=0.73,\te=19)$}
\end{figure}
\begin{figure}[H]
\centerline{
\includegraphics[width=12.5cm, height=6.75cm,
trim=1.5cm 7cm 2cm 6.7cm,clip
]{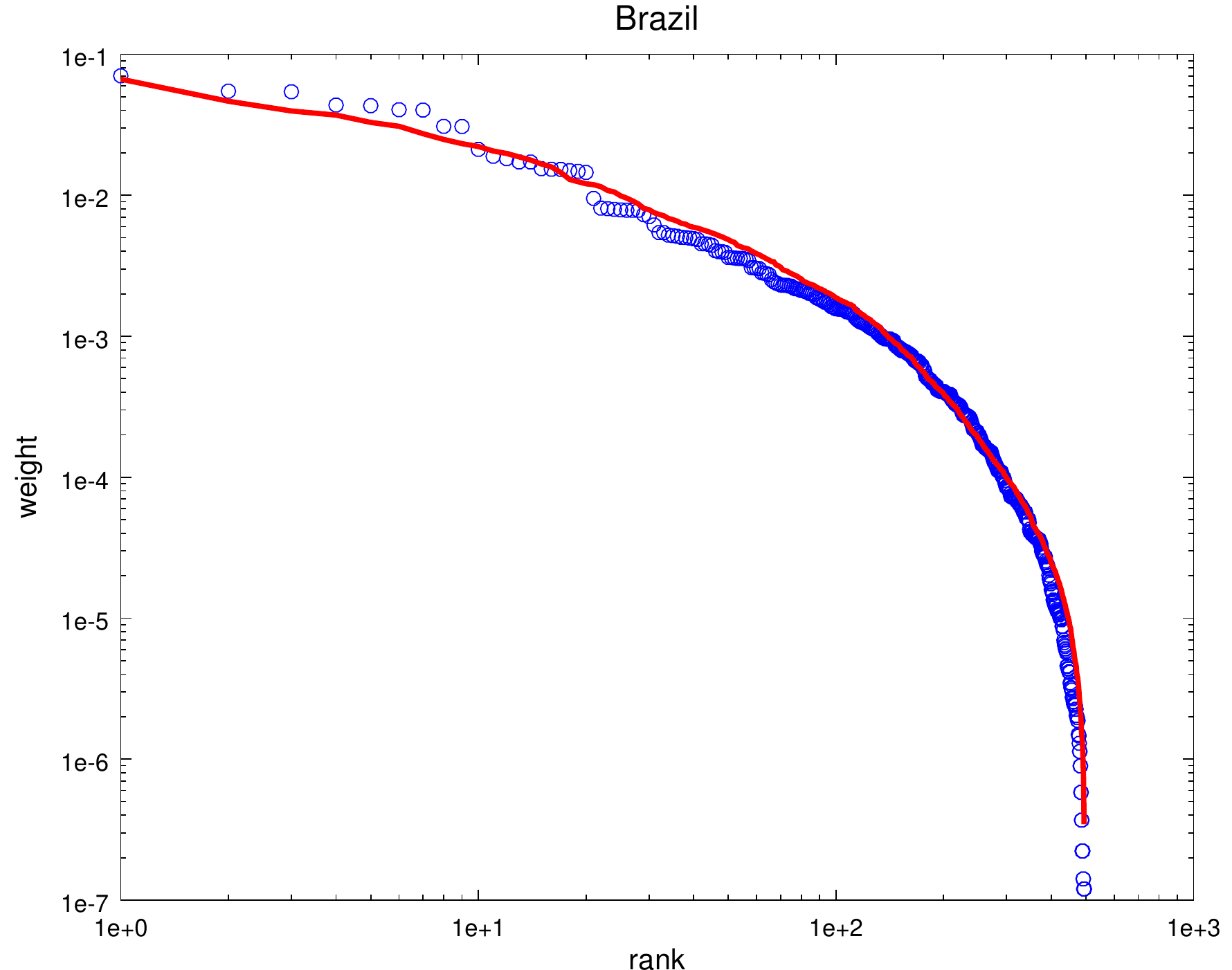}}
\caption{\small Brazil $(\al=0.1,\te=50)$}
\end{figure}

\begin{figure}[H]
\centerline{
\includegraphics[width=12.5cm, height=6.75cm,
trim=1.5cm 7cm 2cm 6.7cm,clip
]{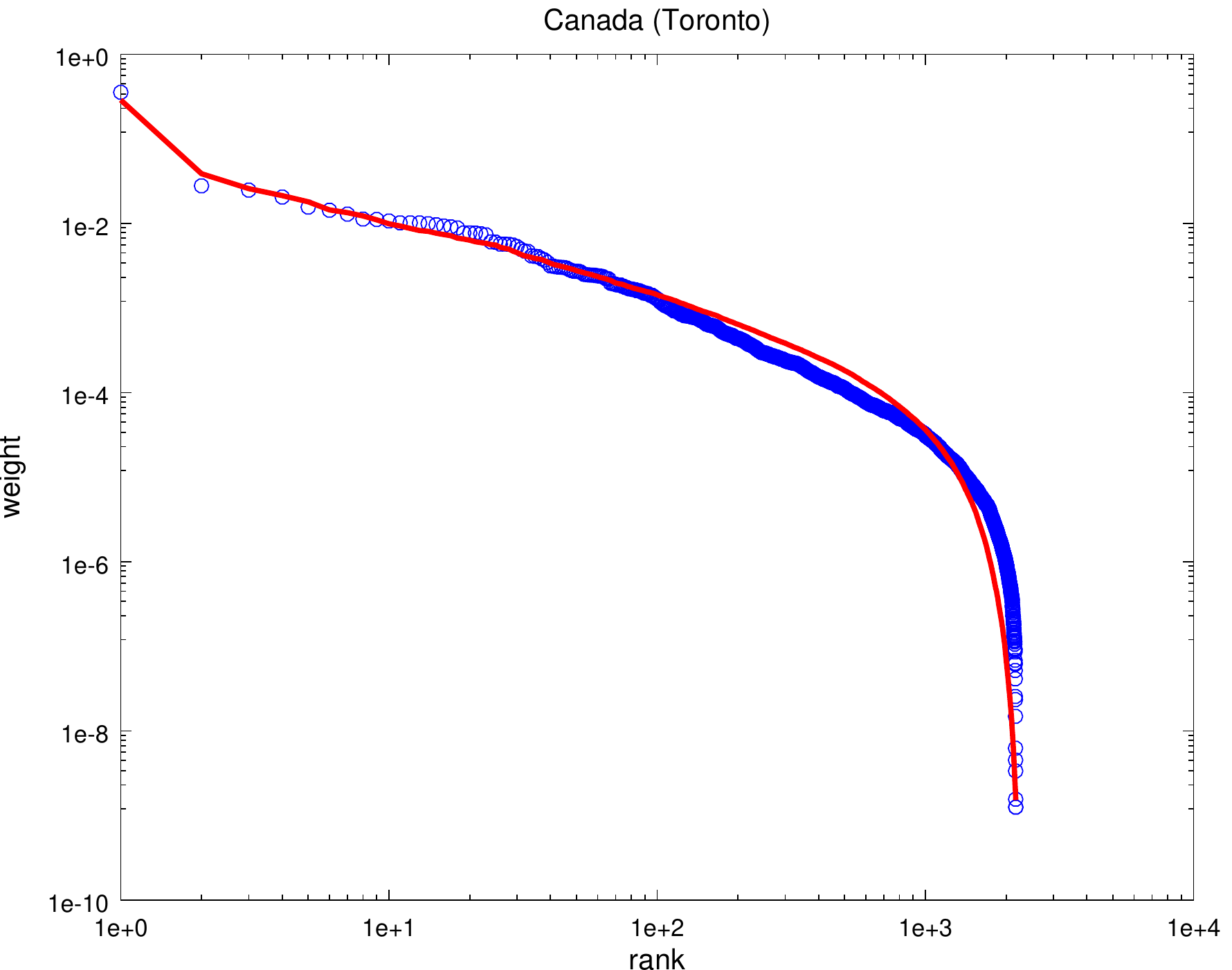}}
\caption{\small Canada (w/o {\sc CNOOC}) $(\al=0.75,\te=40)$}
\end{figure}

\begin{figure}[H]
\centerline{
\includegraphics[width=12.5cm, height=6.75cm,
trim=1.5cm 7cm 2cm 6.7cm,clip
]{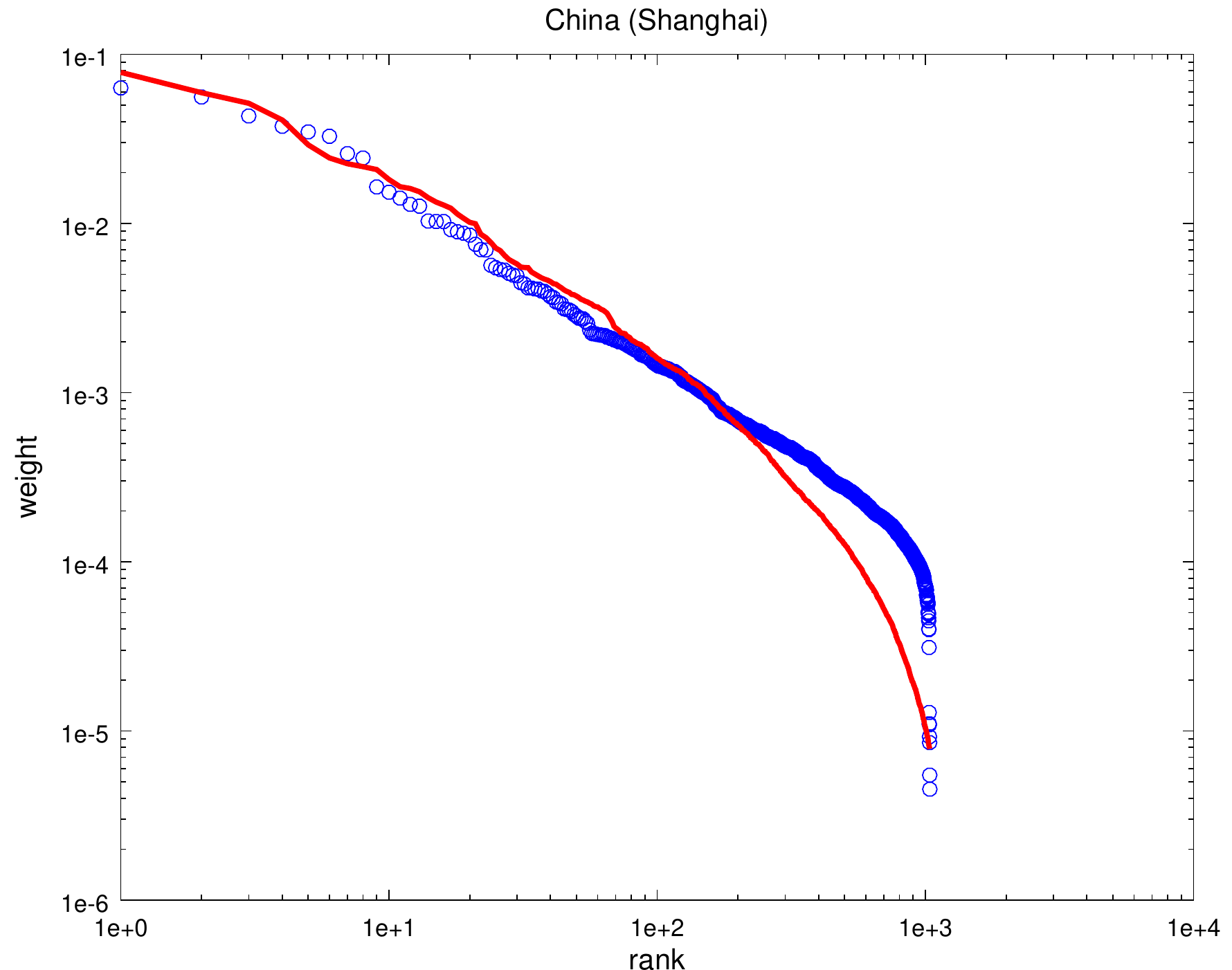}}
\caption{\small China (Shanghai) $(\al=0.57,\te=20)$}
\end{figure}

\begin{figure}[H]
\centerline{
\includegraphics[width=12.5cm, height=6.75cm,
trim=1.5cm 7cm 2cm 6.7cm,clip
]{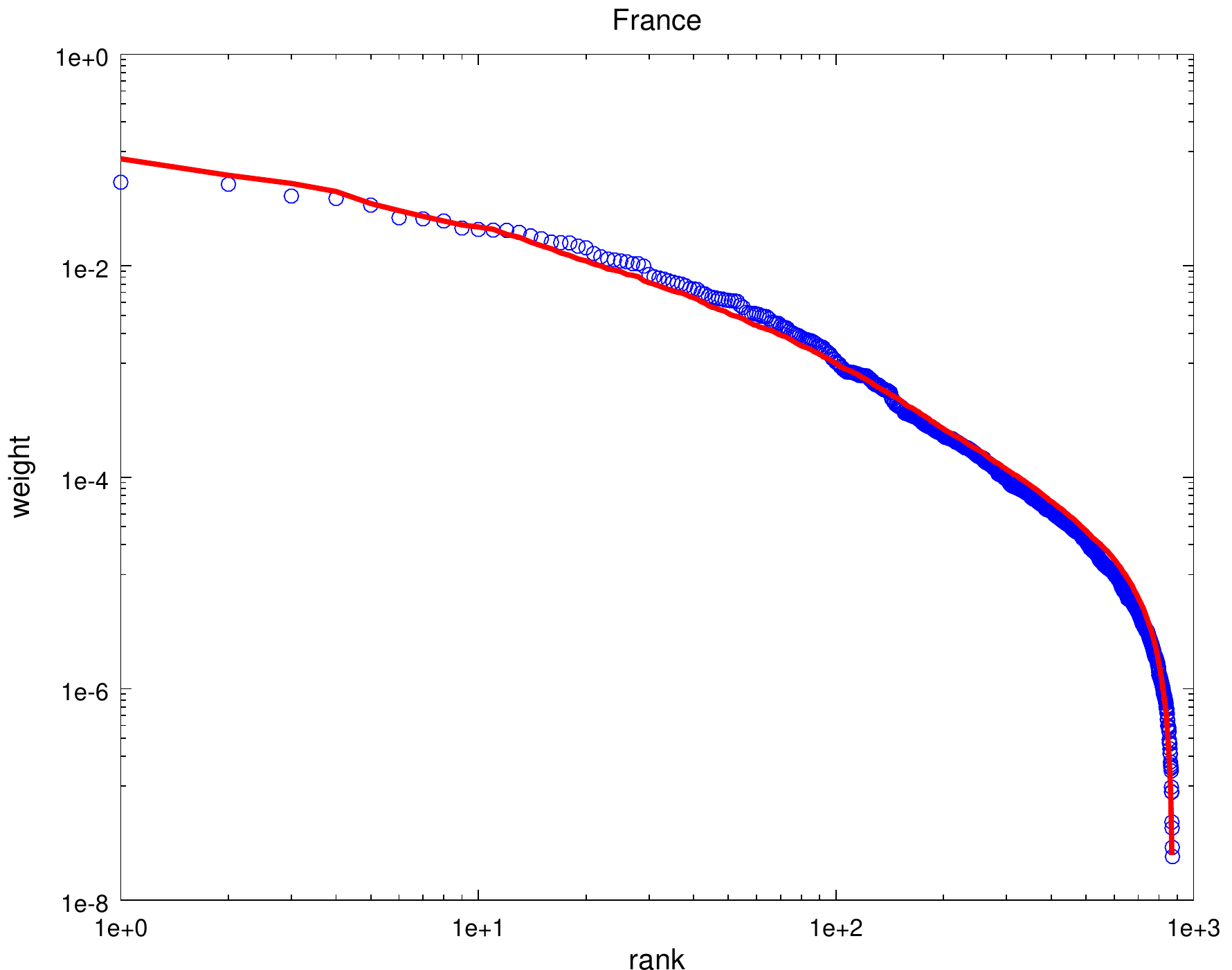}}
\caption{\small France $(\al=0.35,\te=20)$}
\end{figure}

\begin{figure}[H]
\centerline{
\includegraphics[width=12.5cm, height=6.75cm,
trim=1.5cm 7cm 2cm 6.7cm,clip
]{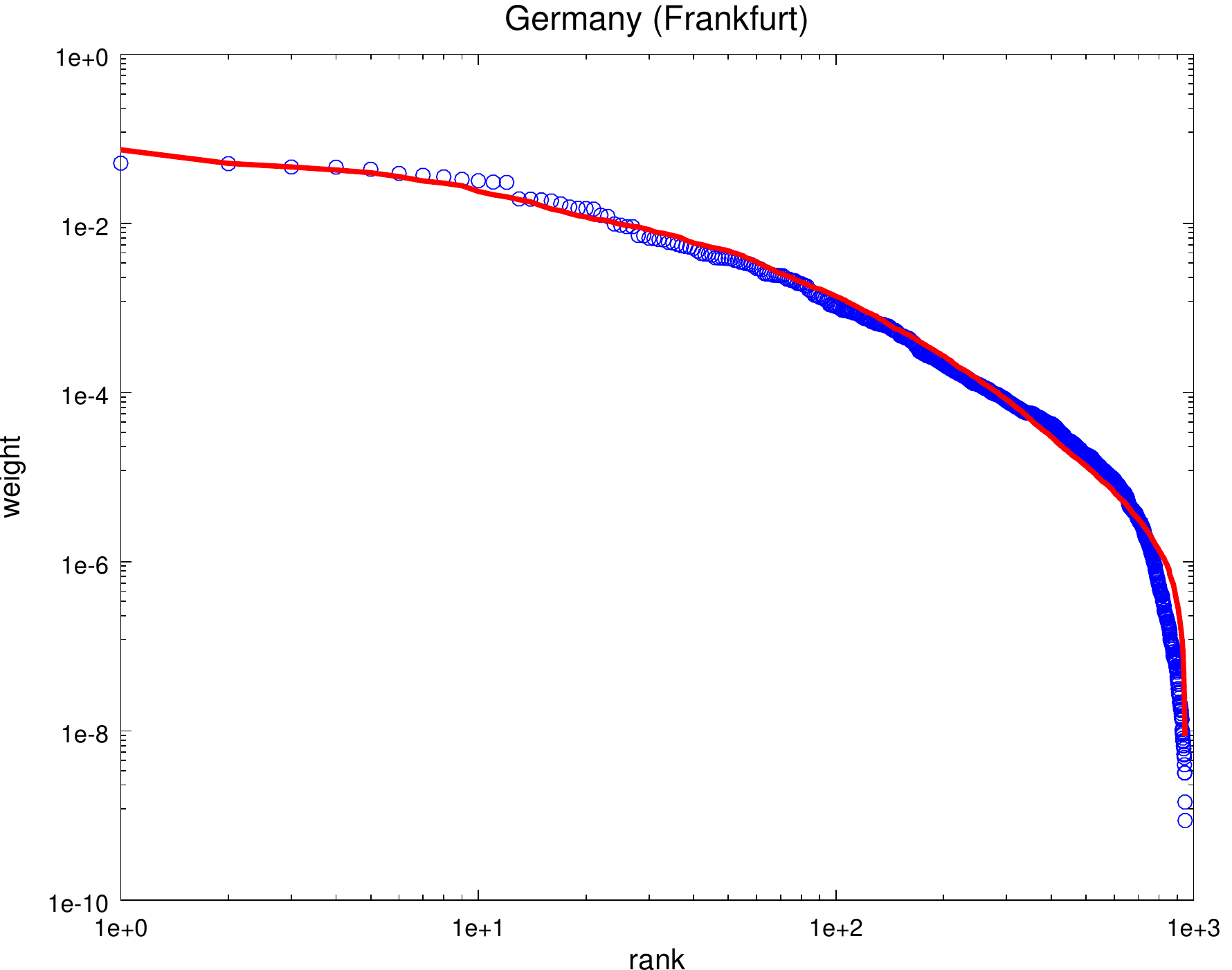}}
\caption{\small Germany $(\al=0.2,\te=34)$}
\end{figure}

\begin{figure}[H]
\centerline{
\includegraphics[width=12.5cm, height=6.75cm,
trim=1.5cm 7cm 2cm 6.7cm,clip
]{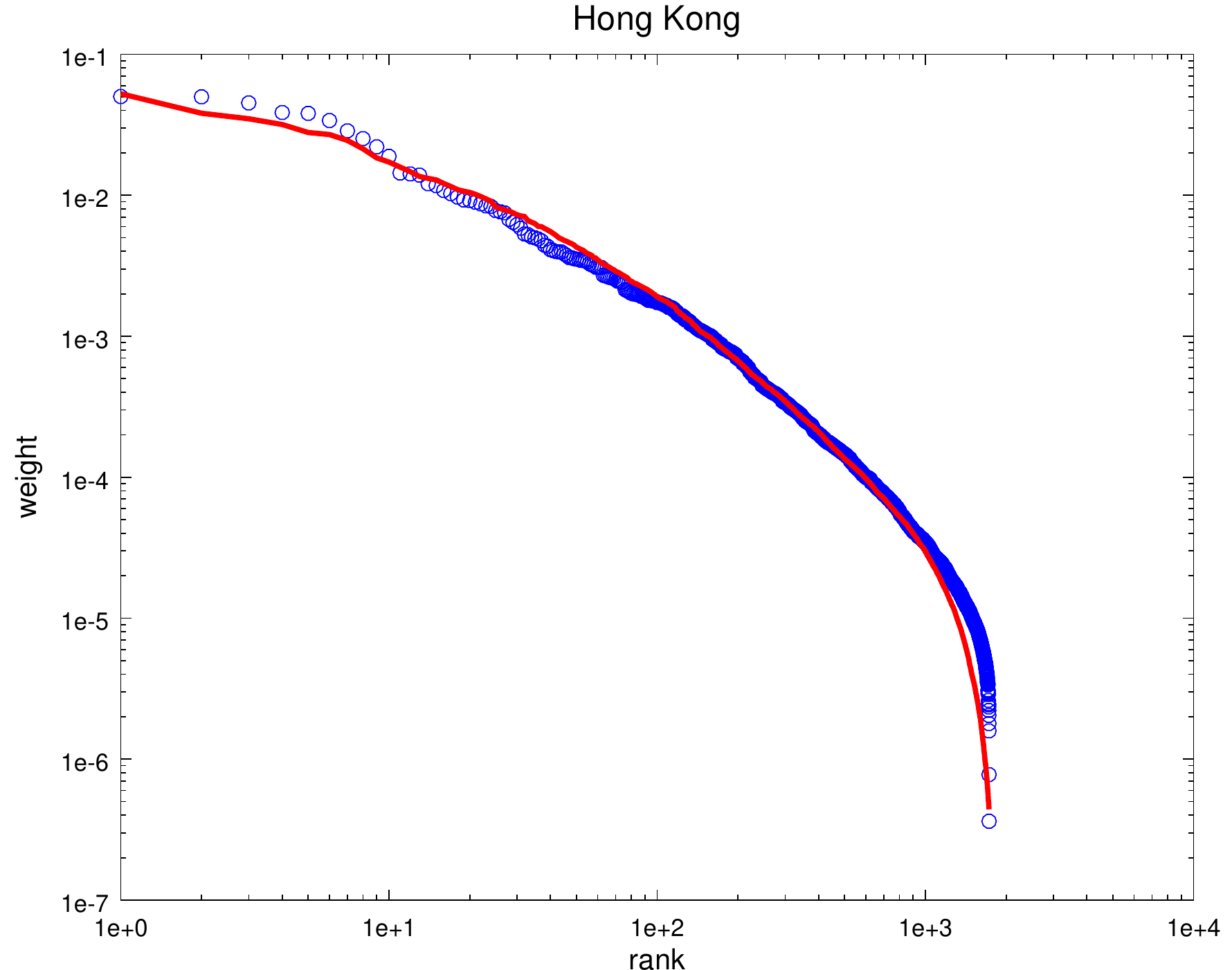}}
\caption{\small Hong Kong $(\al=0.47,\te=40)$}
\end{figure}
\begin{figure}[H]
\centerline{
\includegraphics[width=12.5cm, height=6.75cm,
trim=1.5cm 7cm 2cm 6.7cm,clip
]{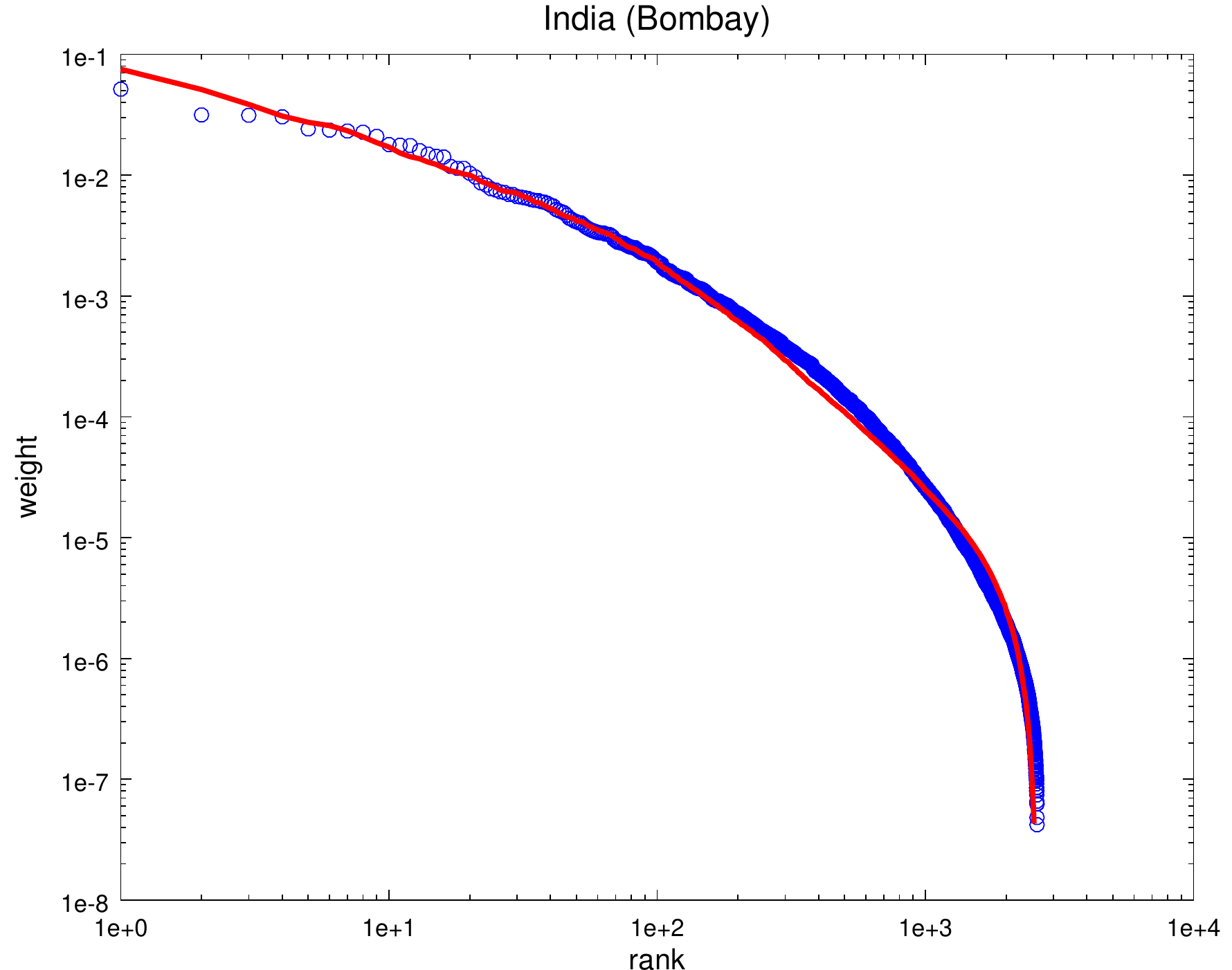}}
\caption{\small India $(\al=0.43,\te=37)$}
\end{figure}

\begin{figure}[H]
\centerline{
\includegraphics[width=12.5cm, height=6.75cm,
trim=1.5cm 7cm 2cm 6.7cm,clip
]{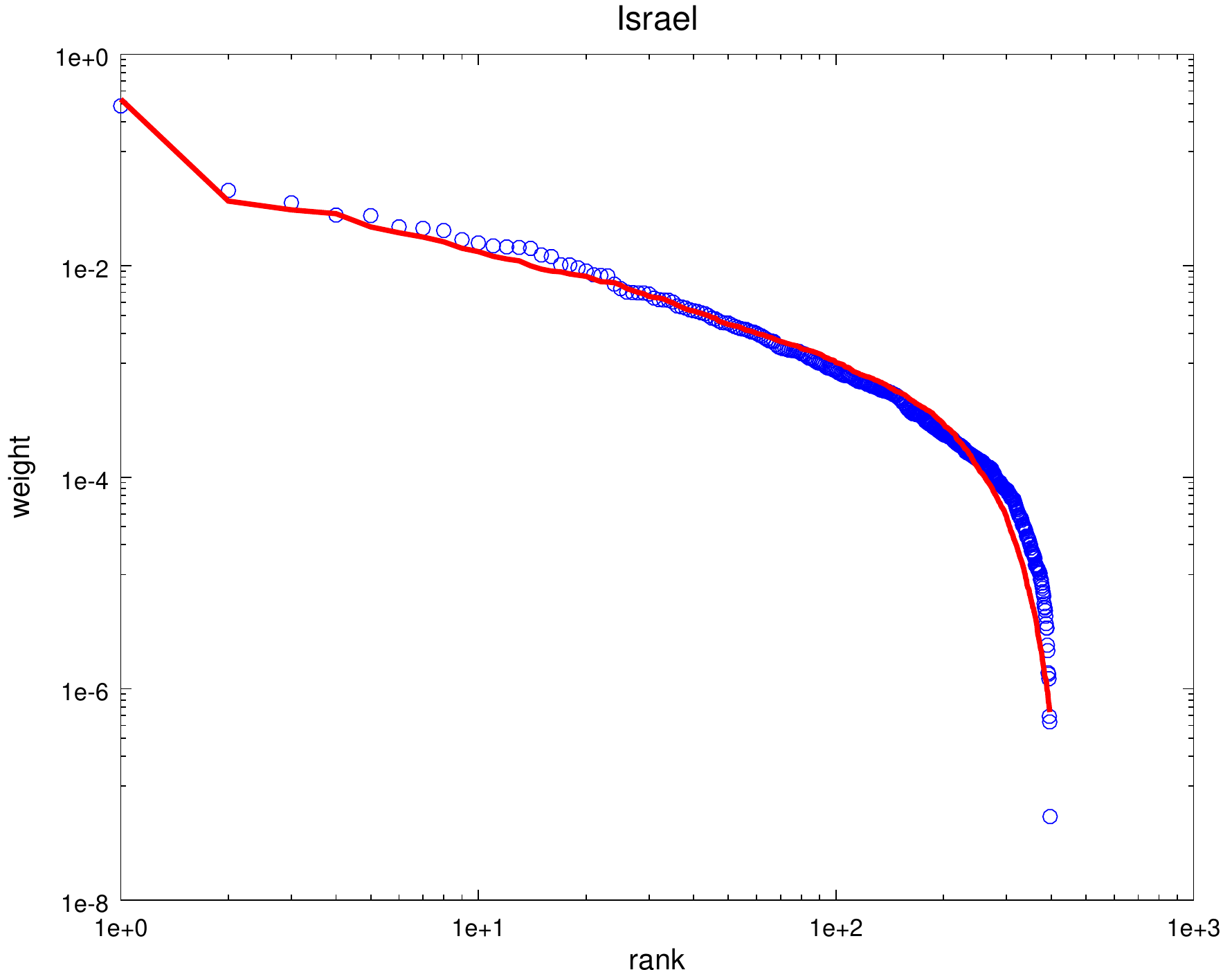}}
\caption{\small Israel $(\al=0.65,\te=45)$}
\end{figure}
\begin{figure}[H]
\centerline{
\includegraphics[width=12.5cm, height=6.75cm,
trim=1.5cm 7cm 2cm 6.7cm,clip
]{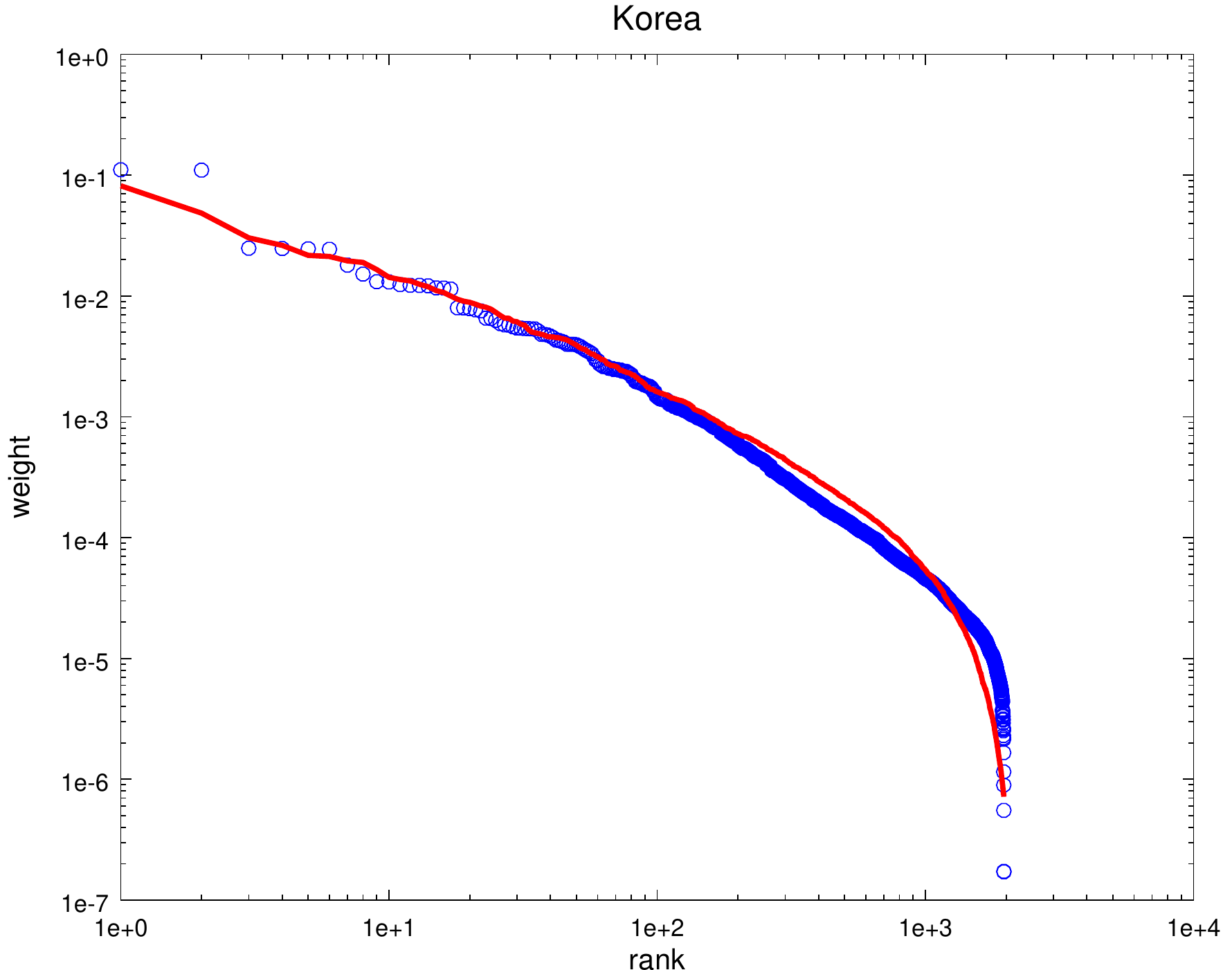}}
\caption{\small Korea $(\al=0.55,\te=45)$}
\end{figure}

\begin{figure}[H]
\centerline{
\includegraphics[width=12.5cm, height=6.75cm,
trim=1.5cm 7cm 2cm 6.7cm,clip
]{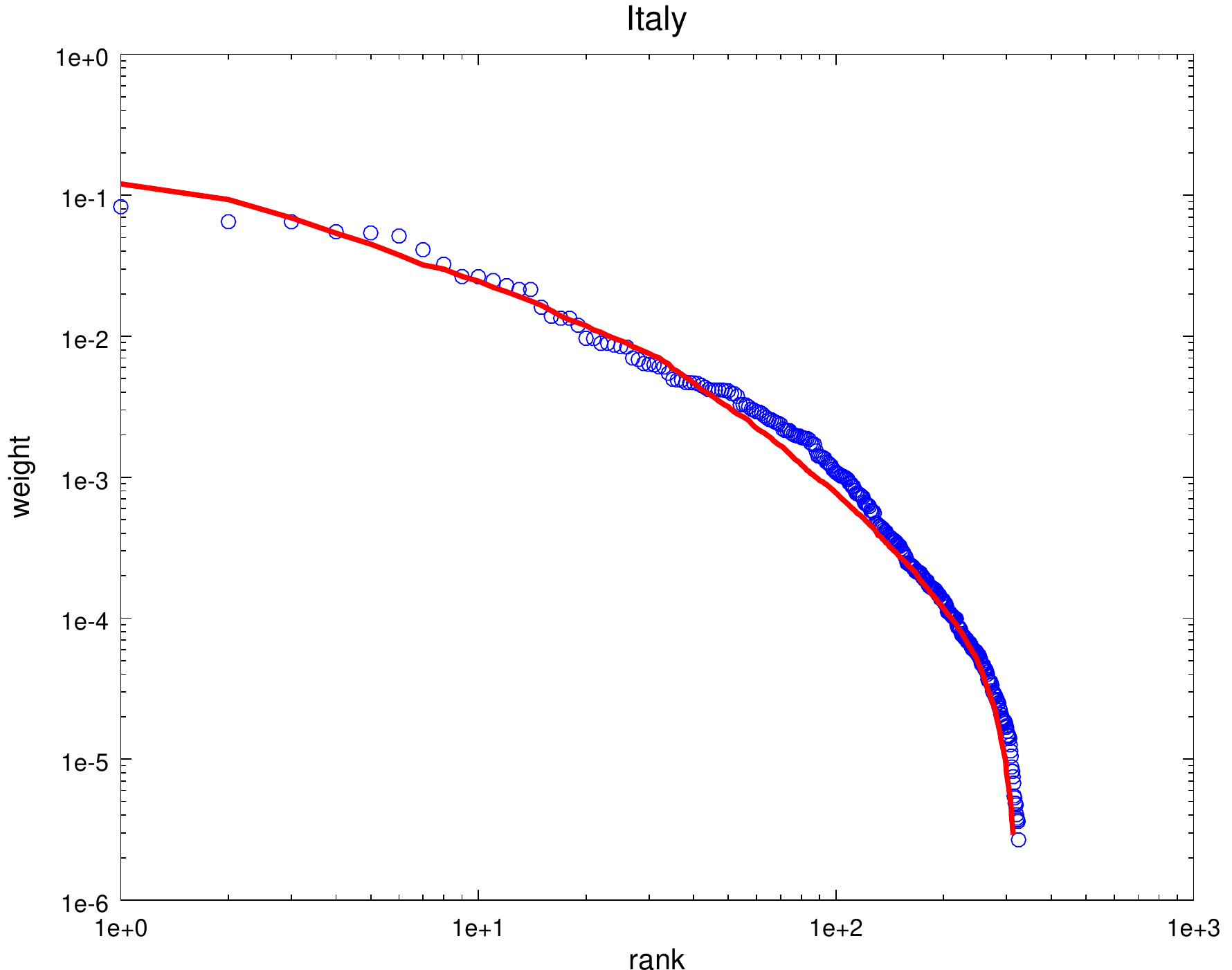}}
\caption{\small Italy $(\al=0.3,\te=15)$}
\end{figure}

\begin{figure}[H]
\centerline{
\includegraphics[width=12.5cm, height=6.75cm,
trim=1.5cm 7cm 2cm 6.7cm,clip
]{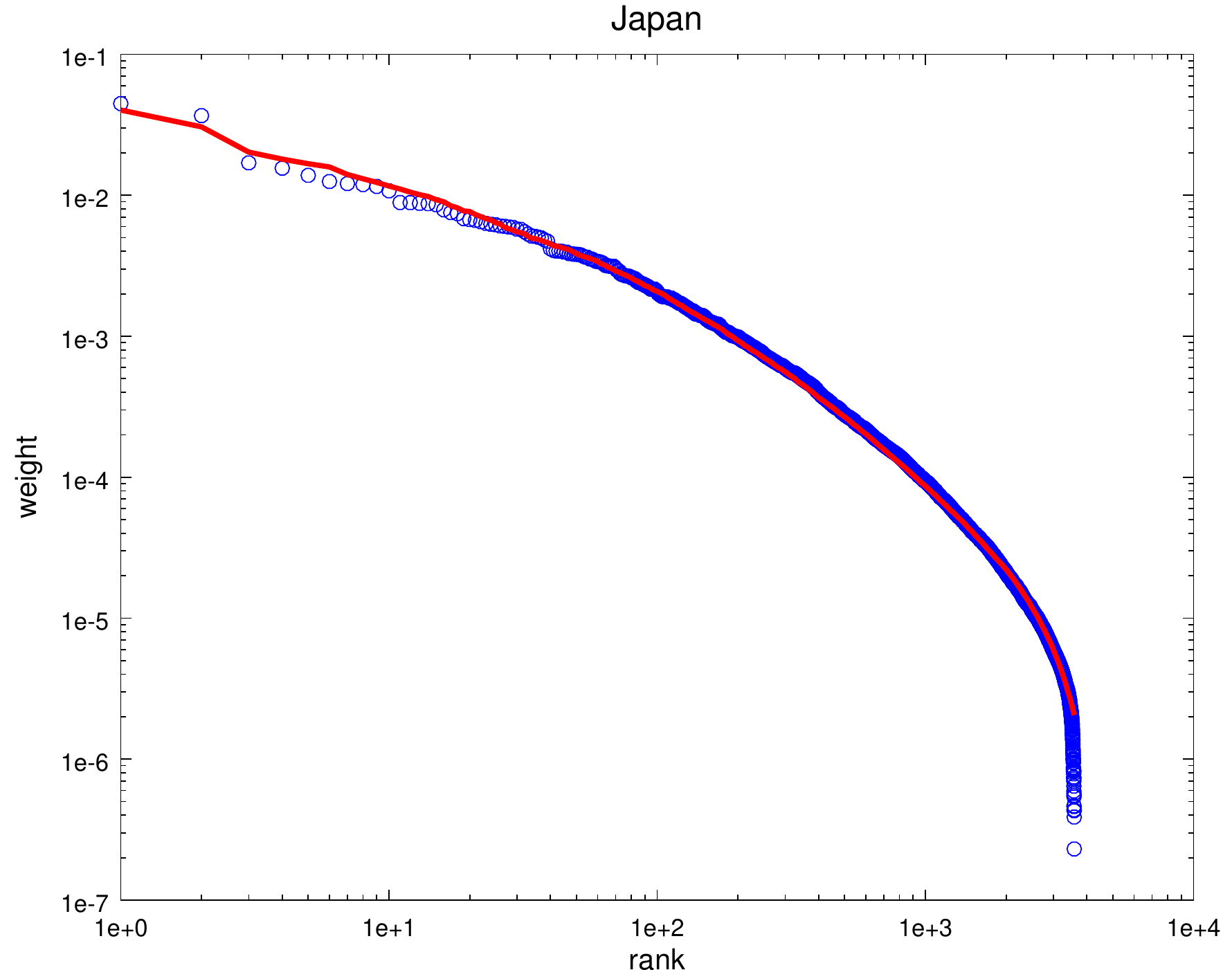}}
\caption{\small Japan $(\al=0.48,\te=95)$}
\end{figure}
\begin{figure}[H]
\centerline{
\includegraphics[width=12.5cm, height=6.75cm,
trim=1.5cm 7cm 2cm 6.7cm,clip
]{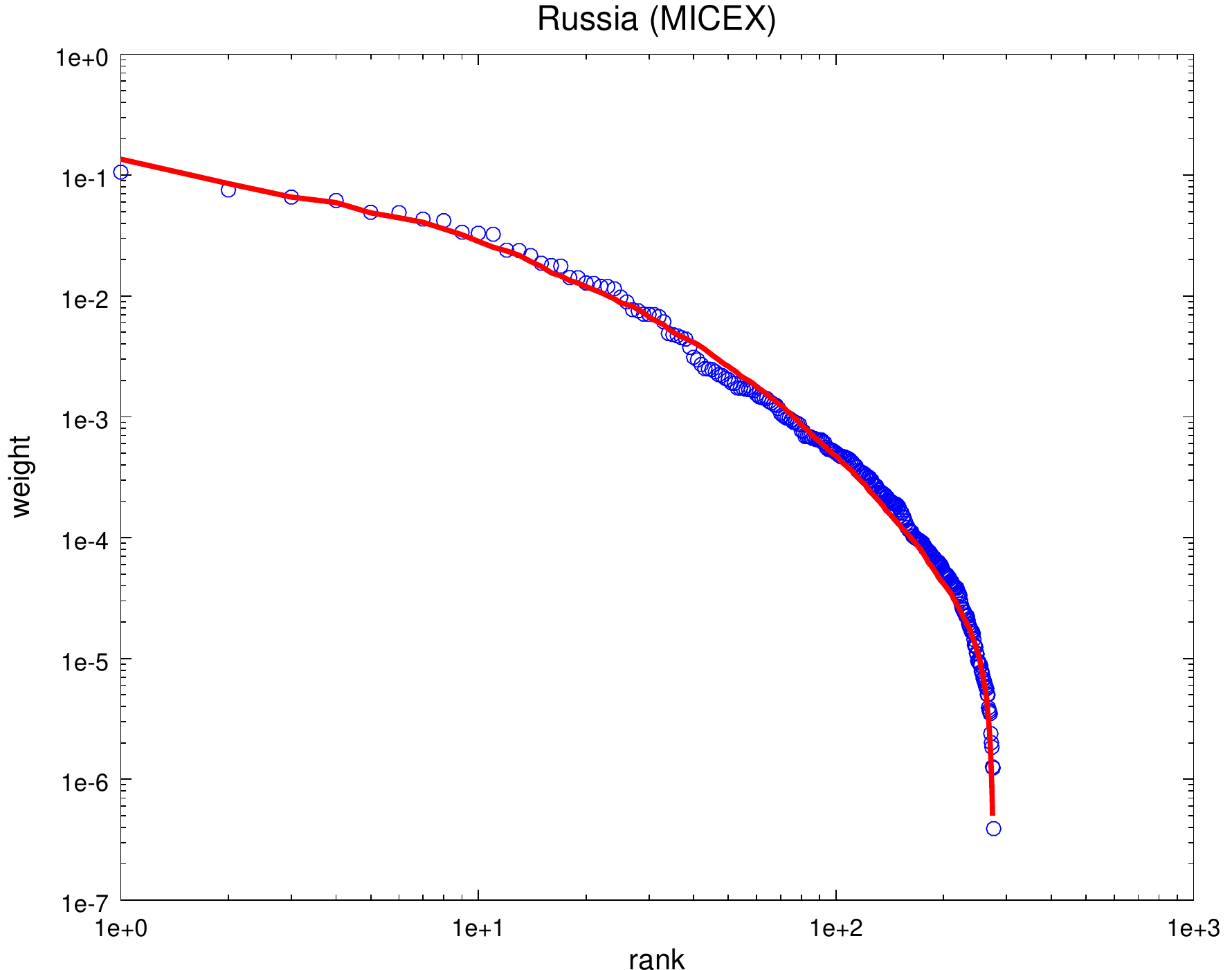}}
\caption{\small Russia $(\al=0.2,\te=15)$}
\end{figure}

\begin{figure}[H]
\centerline{
\includegraphics[width=12.5cm, height=6.75cm,
trim=1.5cm 7cm 2cm 6.7cm,clip
]{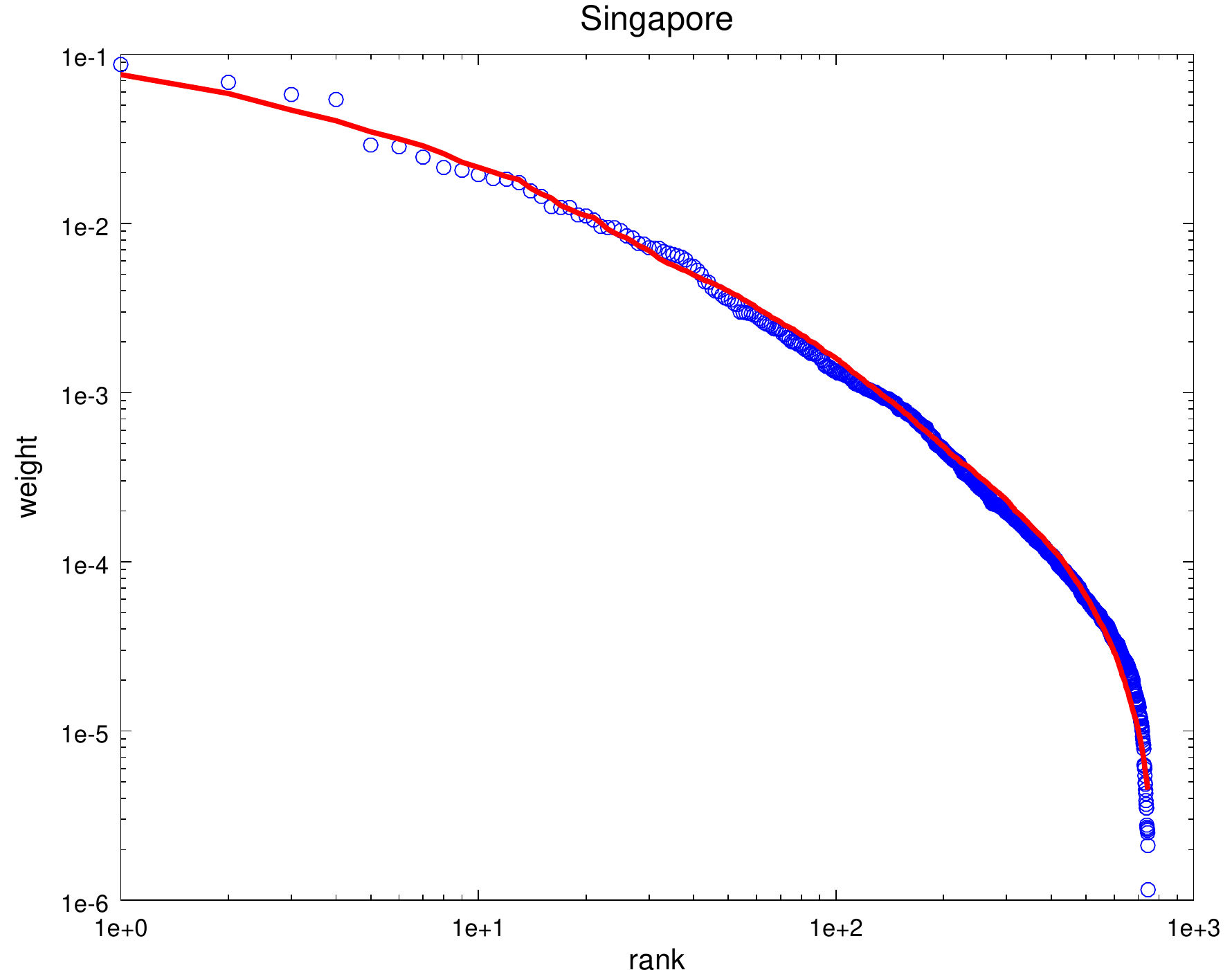}}
\caption{\small Singapore $(\al=0.51,\te=20)$}
\end{figure}

\begin{figure}[H]
\centerline{
\includegraphics[width=12.5cm, height=6.75cm,
trim=1.5cm 7cm 2cm 6.7cm,clip
]{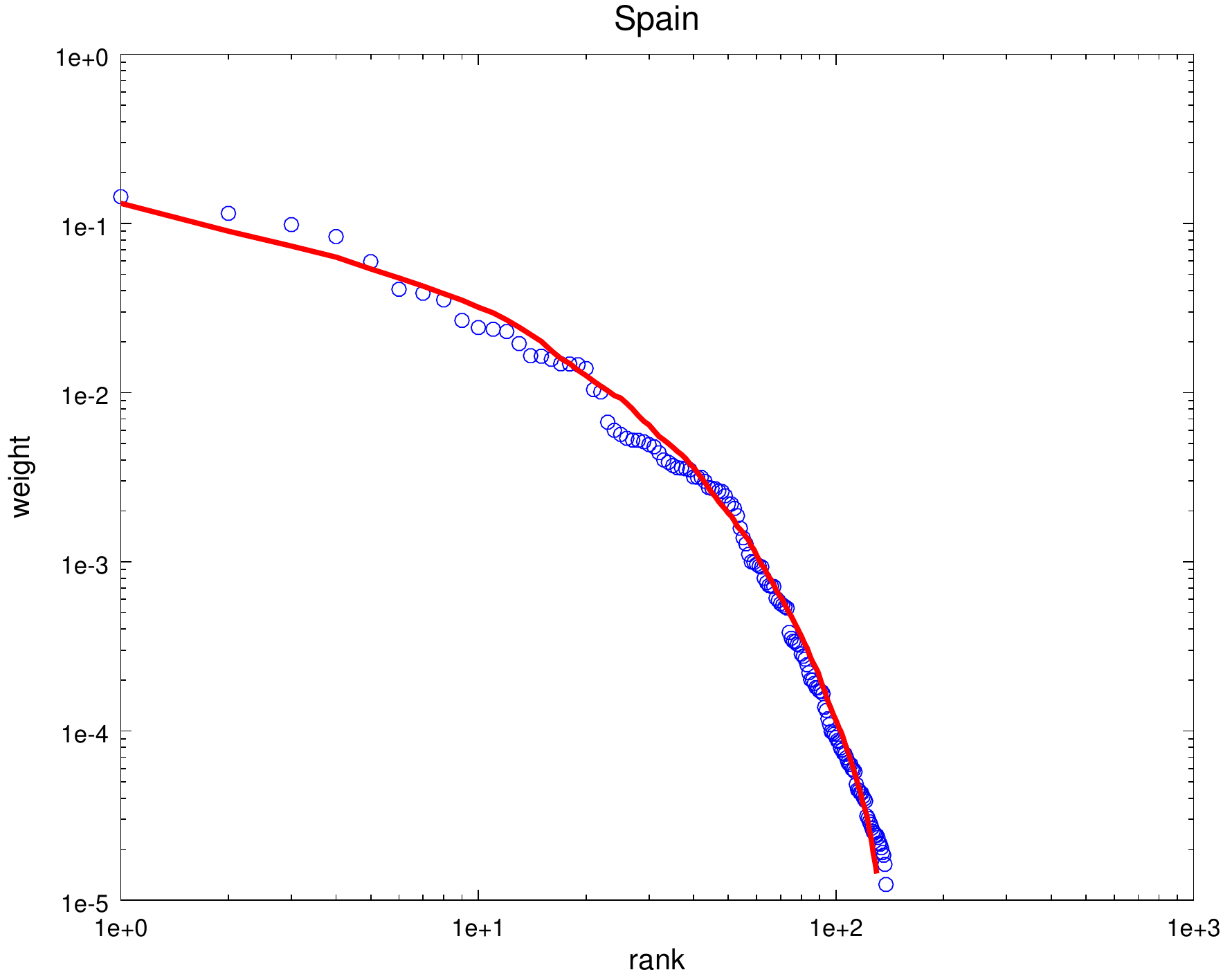}}
\caption{\small Spain $(\al=0.02,\te=16)$}
\end{figure}
\begin{figure}[H]
\centerline{
\includegraphics[width=12.5cm, height=6.75cm,
trim=1.5cm 7cm 2cm 6.7cm,clip
]{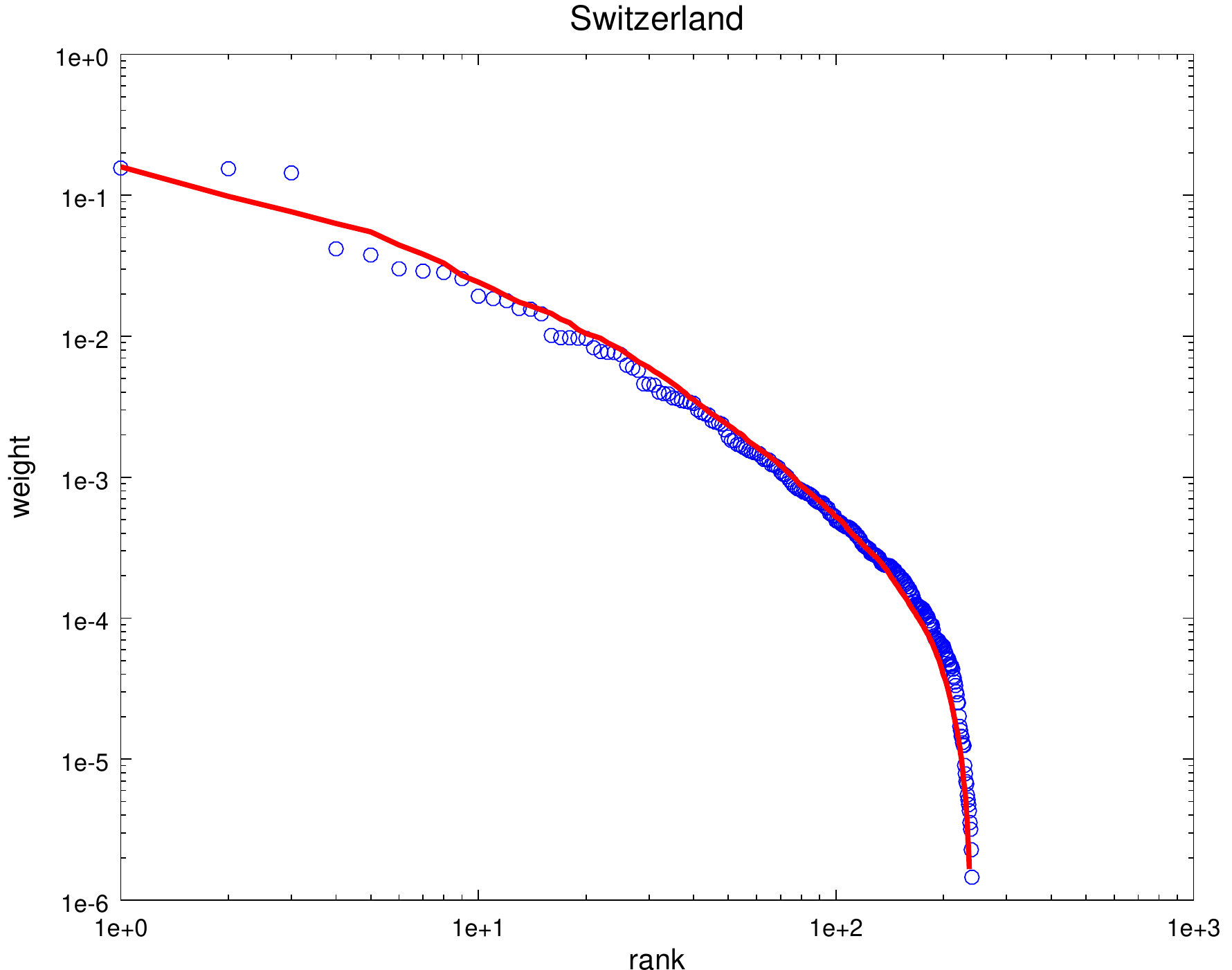}}
\caption{\small Switzerland $(\al=0.4,\te=9)$}
\end{figure}

\begin{figure}[H]
\centerline{
\includegraphics[width=12.5cm, height=6.75cm,
trim=1.5cm 7cm 2cm 6.7cm,clip
]{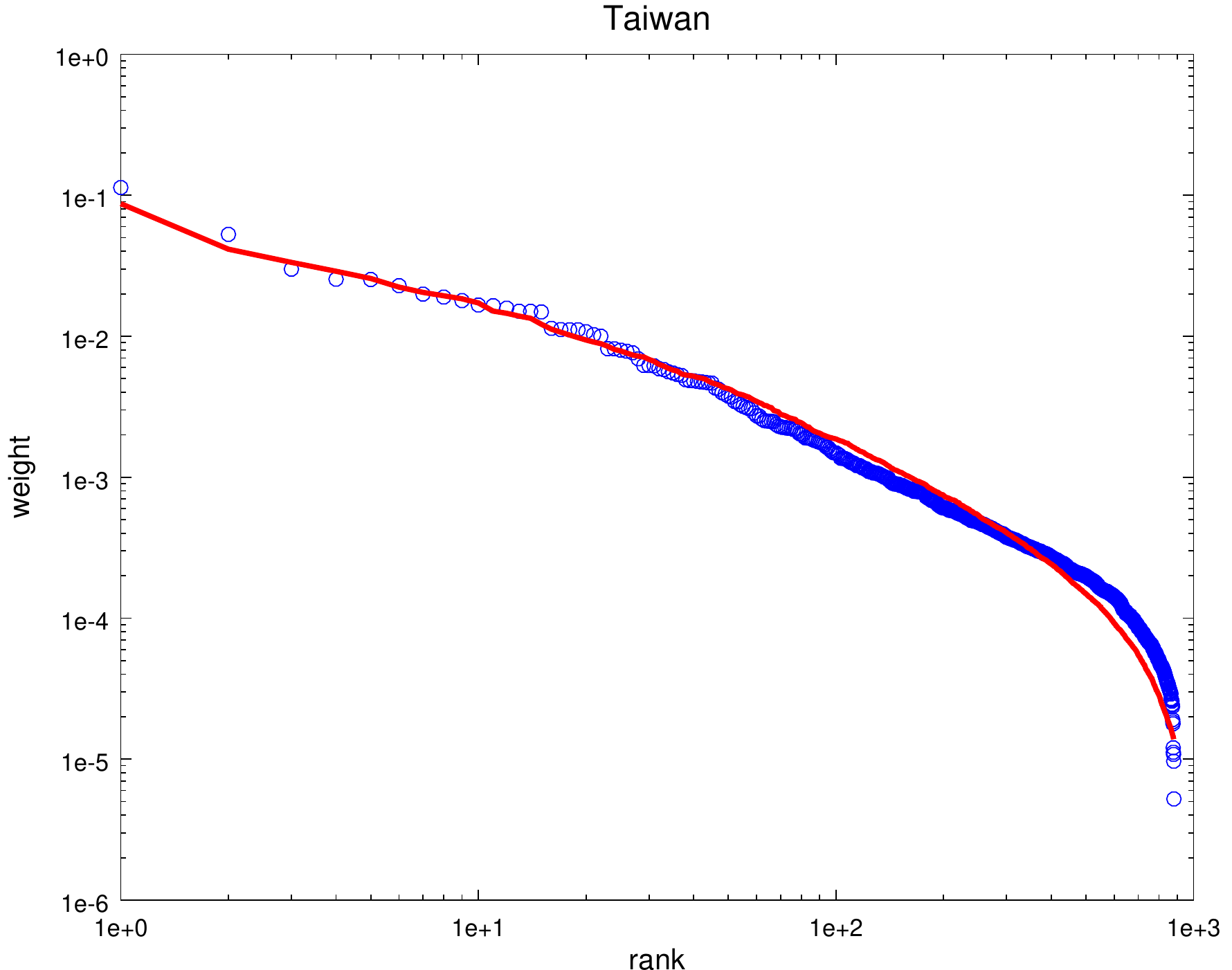}}
\caption{\small Taiwan $(\al=0.5,\te=50)$}
\end{figure}

\begin{figure}[H]
\centerline{
\includegraphics[width=12.5cm, height=6.75cm,
trim=1.5cm 7cm 2cm 6.7cm,clip
]{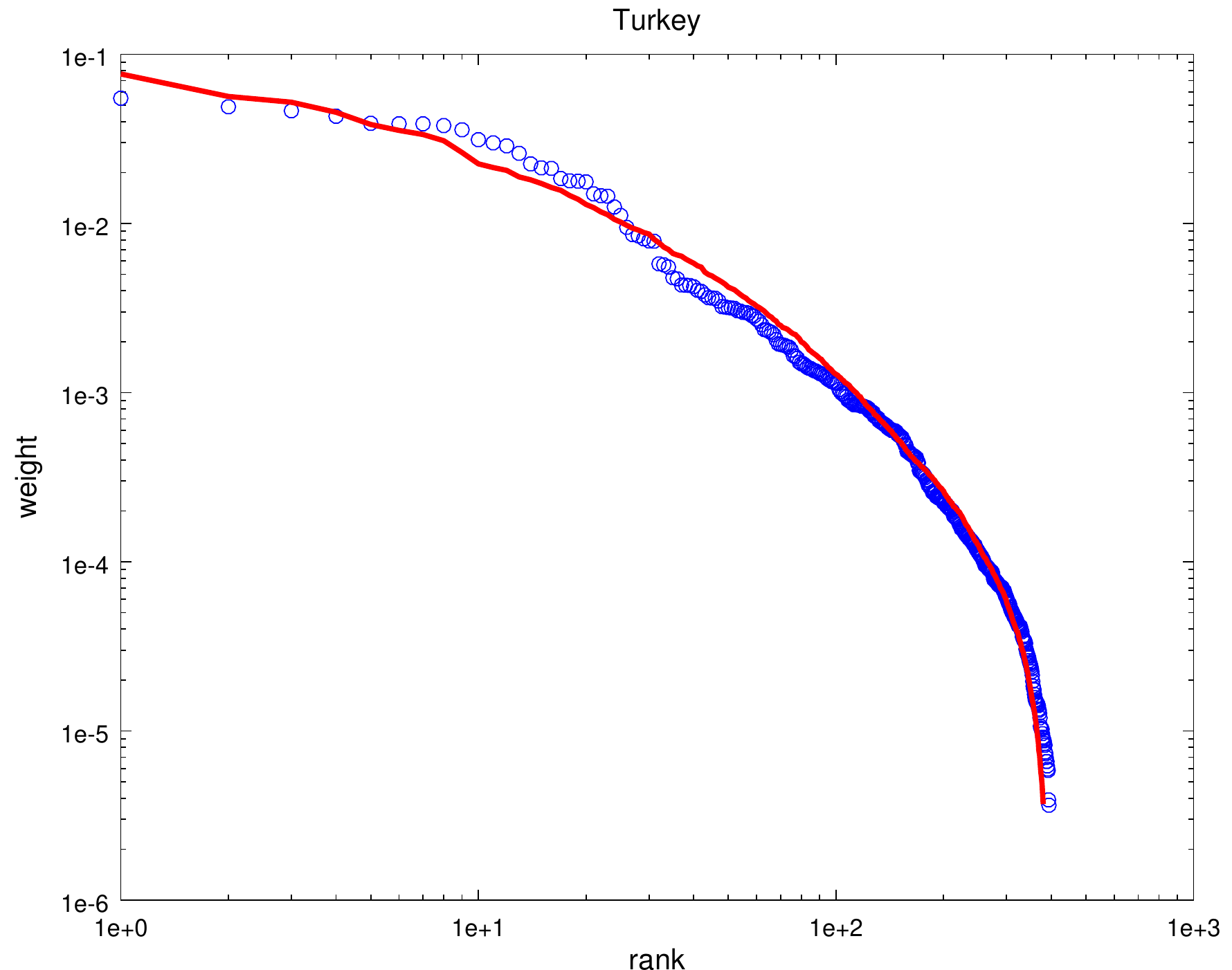}}
\caption{\small Turkey $(\al=0.24,\te=26)$}
\end{figure}

\begin{figure}[H]
\centerline{
\includegraphics[width=12.5cm, height=7cm,
trim=1.5cm 7cm 2cm 6.7cm,clip
]{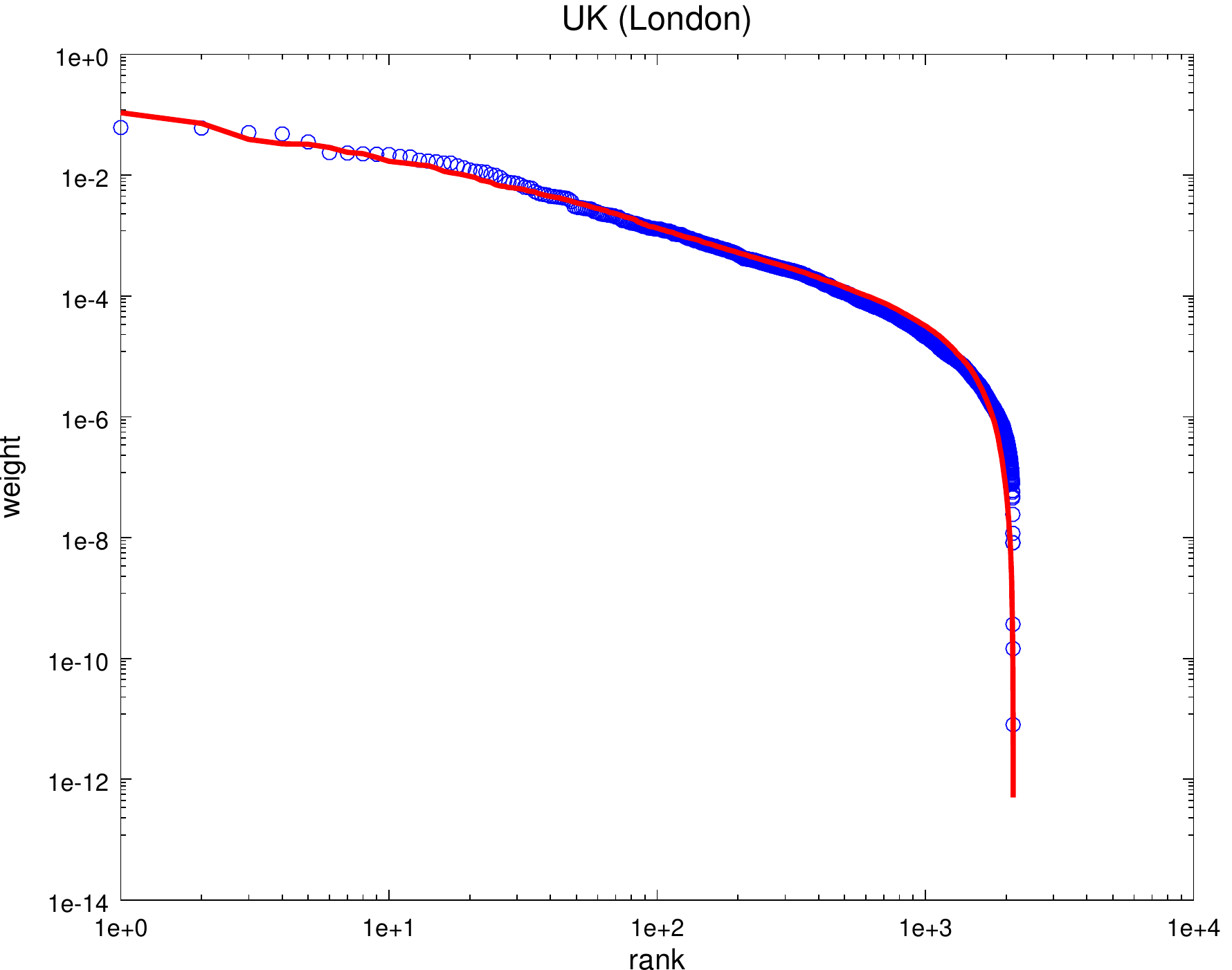}}
\caption{\small United Kingdom  $(\al=0.65,\te=15)$}
\end{figure}

\begin{figure}[H]
\centerline{
\includegraphics[width=12.5cm, height=6.75cm,
trim=1.5cm 7cm 2cm 6.7cm,clip]{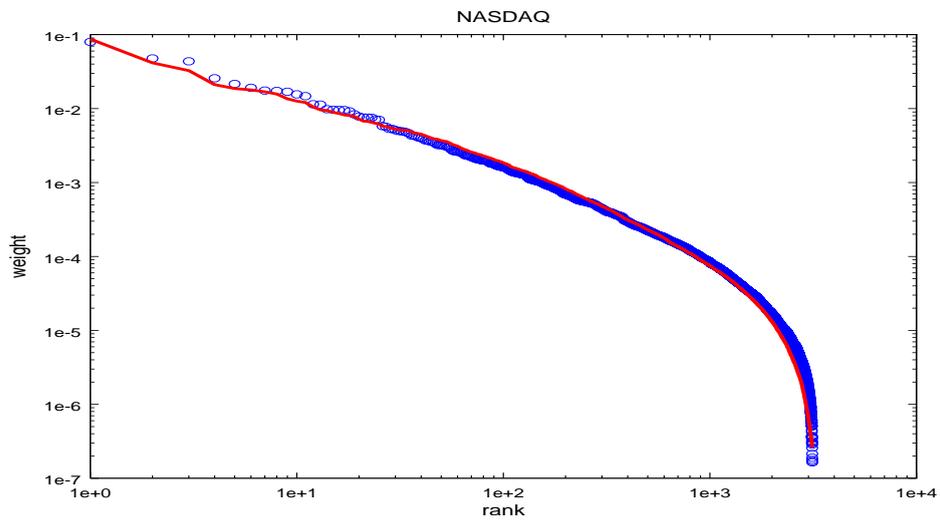}}
\caption{\small United States $(\al=0.60,\te=55)$}
\end{figure}

\begin{figure}[H]
\centerline{
\includegraphics[width=12.5cm, height=6.75cm,
trim=1.5cm 7cm 2cm 6.7cm,clip]{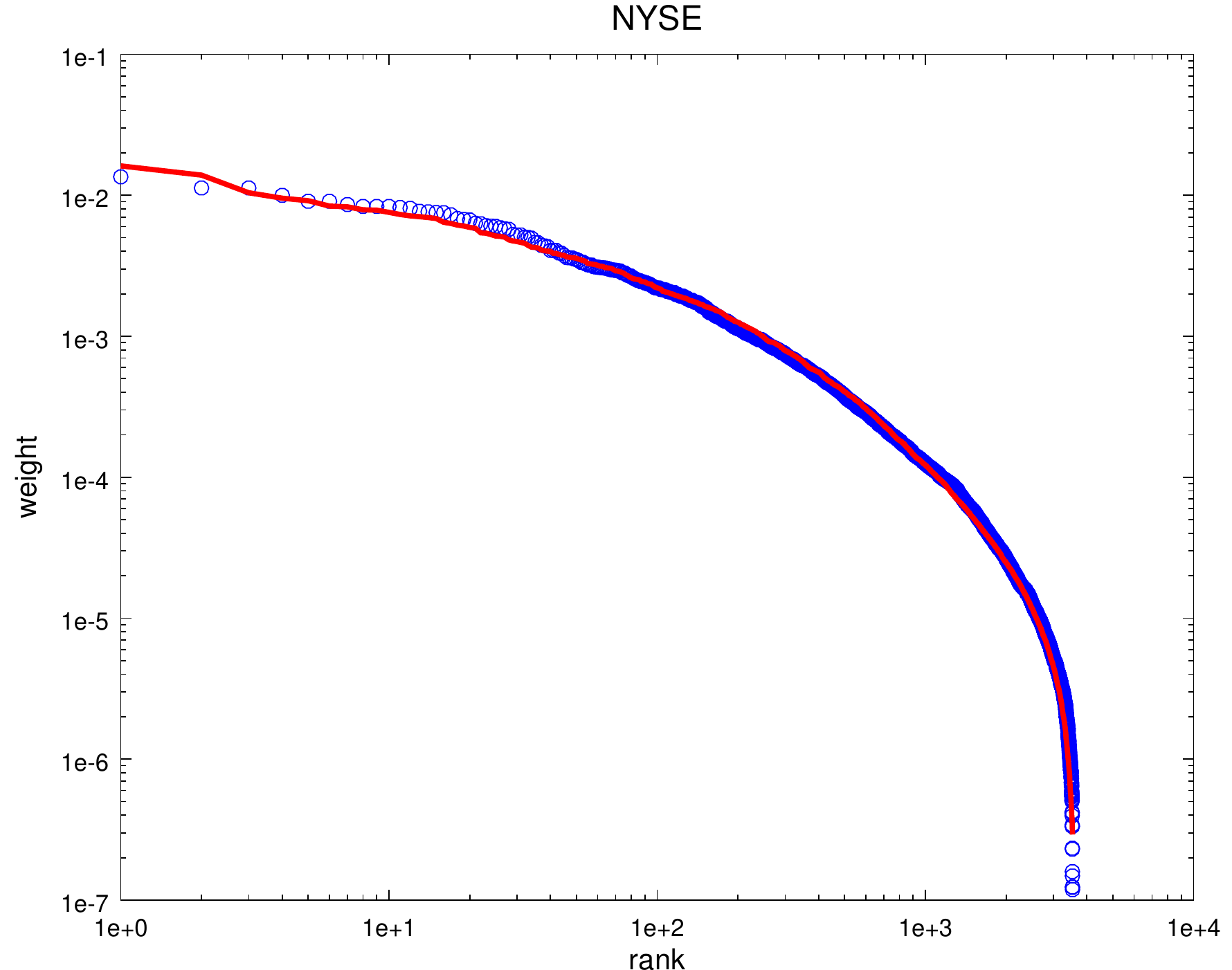}}
\caption{\small United States $(\al=0.28,\te=255)$}
\end{figure}


\newpage
\subsection{S\&P 500 drawdowns}\label{subsec-drawdown}
Figure in this section illustrates another application of the two-parameter \PD distribution providing fit to normalized relative daily drawdowns of S\&P 500.
\begin{figure}[H]
\centerline{
\includegraphics[width=14cm, height=7.5cm,
]{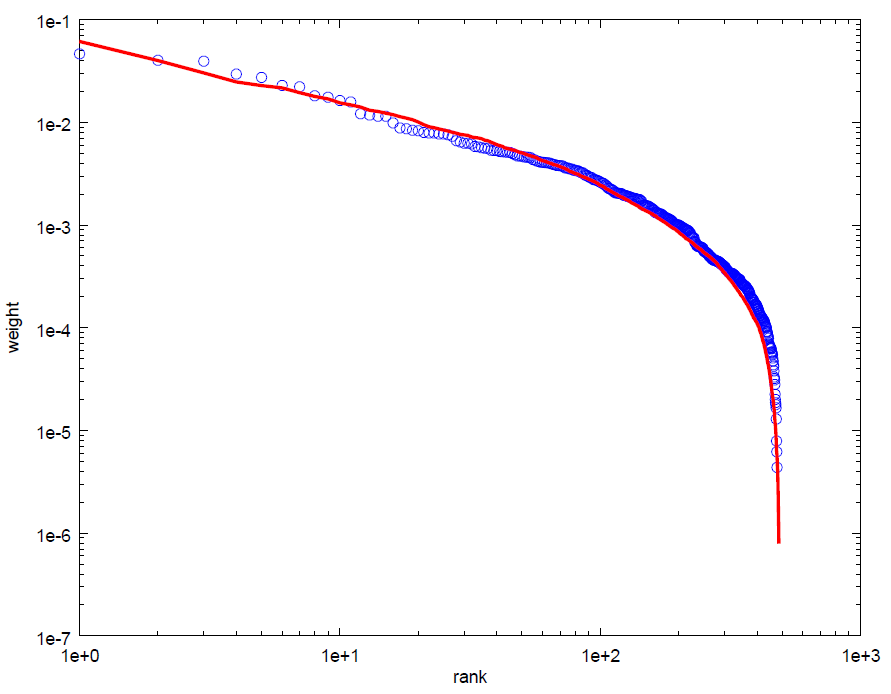}}
\caption{\small S\&P 500 relative daily drawdowns, from 1950 to 2014; $\te=80,\al=0.24$}
\end{figure}

\section{Summary}
The central idea of the proposed approach is probabilistic-combinatorial and
can be summarized as follows:
\begin{itemize}
	\item {\cldb \tt combinatorics:} Stock market is considered as large combinatorial structure - partition of the set of all invested units of money. Stock capitalizations, represented by integers, define partitions of the market value. Number of ways
this state can be realized combinatorially is given by the formula \eqref{numc}.
	\item {\cldb \tt probability:} Partition structure for each level $n\ge1$ defines \exc probability for all partitions with $n$ elements, such that 
 distribution on level $n$ is consistent with distribution on partitions with $n+1$ elements. Such consistency conditions determine up ($n\to n+1$) and down ($n\to n-1$) conditional probabilities of transitions.
	\item {\cldb The \tpPD distribution}, defined in the infinite simplex with ranked weights has corresponding partition structure, given by the formula \eqref{PSF}.
Stick-breaking construction provides size-biased method of sampling from the two-parameter model. Associated diffusion process, induced by down/up Markov chains has $\PDat$ law as unique reversible and therefore equilibrium distribution.
\end{itemize}

Results of Section \ref{sec-examples} suggest the hypothesis that the two-parameter model approximates stationary distribution of capital distribution curve as well as  corresponding underlying partition structure. It is proposed that vector of the ranked market weights (capital distribution curve) fluctuates in stochastic equilibrium, which can be modelled by means of the two-parameter diffusion process, or by combinatorial random walks on partitions. 

\newpage
\bibliographystyle{abbrv}
\bibliography{REFF}

\end{document}

%% file: SDefin.tex
\usepackage{xstring}
\usepackage{xifthen}

\usepackage{color}
\newcommand{\clr}{\color{red}}
\newcommand{\clb}{\color{blue}}
\newcommand{\cldb}{\color{blue!50!black}}

\renewcommand{\ge}{\geqslant}
\renewcommand{\le}{\leqslant}
\newcommand{\te}{\theta}

\newcommand{\eqn}{\begin{equation}}
\newcommand{\nqe}{\end{equation}}

\newcommand{\vc}[1]{\boldsymbol{\mathsf #1}}
\newcommand{\sfp}[1]{  \boldsymbol{\mathsf p}_{\mathsf #1}}
\newcommand{\sfpi}[1]{  \boldsymbol{\mathsf \pi}_{\mathsf #1}}

\newcommand{\EE}{\mathbb{E}}
\newcommand{\RR}{\mathbb{R}}

\newcommand{\Bd}{\mathcal{B}}
\newcommand{\Gd}{\mathcal{G}}
\newcommand{\GG}{\Gamma}
\newcommand{\BB}{\text{B}}
\newcommand{\Dir}{\mathcal{D}}

\newcommand{\PD}{Poisson-Dirichlet }
\newcommand{\PDat}{\mathcal{PD}({\al,\te})}

\newcommand{\tpPD}{two-parameter Poisson-Dirichlet } 
\newcommand{\prg}[1]{\paragraph{\cldb #1}}

\newcommand{\al}{\alpha}

\newcommand{\la}{\lambda}

\newcommand{\eps}{\varepsilon}

\newcommand*\dif{\mathop{}\!\mathrm{d}}

\newcommand{\vcc}[2]{\vc {#1}=({#1}_1,\dots,{#1}_{#2})}

\newcommand{\Lv}{L\'evy }

\newcommand{\summ}[3]{\sum_{#1=#2}^#3}

\newcommand{\fm}[2]
{
  \StrLen{#2}[\MyStrLen]
  \ifthenelse{\equal{\MyStrLen}{1}}
	{\frac {#1^{#2}}{#2!}}
	{\frac {{#1}^{{#2}}}{({#2})!}}
}

\newcommand{\lif}[2]
{
  \StrLen{#2}[\MyStrLen]
  \ifthenelse{\equal{\MyStrLen}{1}}
	{\frac {#1_#2}{#2!}}
	{\frac {#1_{#2}}{(#2)!}}
}

\newcommand{\rfm}[2]
{
  \StrLen{#1}[\argum]
  \StrLen{#2}[\expon]
	\ifthenelse{\equal{\argum}{1}}
		{\ifthenelse{\equal{\expon}{1}}
			{\frac {#1^{[#2]}}{#2!}}
			{\frac {#1^{[#2]}}{(#2)!}}}
		{\ifthenelse{\equal{\expon}{1}}
			{\frac {(#1)^{[#2]}}{#2!}}
			{\frac {(#1)^{[#2]}}{(#2)!}}}
}

\newcommand{\ffm}[2]
{
  \StrLen{#1}[\argum]
  \StrLen{#2}[\expon]
	\ifthenelse{\equal{\argum}{1}}
		{\ifthenelse{\equal{\expon}{1}}
			{\frac {#1_{(#2)}}{#2!}}
			{\frac {#1_{(#2)}}{(#2)!}}}
		{\ifthenelse{\equal{\expon}{1}}
			{\frac {(#1)_{(#2)}}{#2!}}
			{\frac {(#1)_{(#2)}}{(#2)!}}}
}

\newcommand{\ffl}[2]
{
  \StrLen{#1}[\MyStrLen]
  \ifthenelse{\equal{\MyStrLen}{1}}
	{{#1}_{({#2})}}
	{({#1})_{({#2})}}
}

\newcommand{\rfl}[2]
{
  \StrLen{#1}[\MyStrLen]
  \ifthenelse{\equal{\MyStrLen}{1}}
	{{#1}^{[{#2}]}}
	{({#1})^{[{#2}]}}
}

\newcommand \Ts {\rule{0pt}{2.6ex}}         
\newcommand \Bs {\rule[-0.9ex]{0pt}{0pt}}   
\newcommand \prt {\,\Vdash}
\newcommand \wt {\widetilde}

\newcommand \exc {exchangeable }

\newcommand \pd {$\mathcal{PD}$}

\newcommand \nr {num{\'e}raire }
\newcommand \dn {{\sc dn}}
\newcommand \up {{\sc up}}